\documentclass[12pt]{article}
\usepackage[usenames]{color}
\usepackage{setspace}
\usepackage{graphicx}
\usepackage{amsmath}
\usepackage{amssymb}
\usepackage{amsthm}
\usepackage{multirow}
\numberwithin{equation}{section}

\newcommand{\del}{\partial}

\newcommand{\abs}[1]{\left| #1 \right|}

\newcommand{\bequ}{\begin{align}}
\newcommand{\eequ}{\end{align}}
\newcommand{\beqn}{\begin{align}}
\newcommand{\eeqn}{\end{align}}
\newcommand{\bctr}{\begin{center}}
\newcommand{\ectr}{\end{center}}
\newcommand{\bit}{\begin{itemize}}
\newcommand{\eit}{\end{itemize}}

\newcommand{\half}{{\frac12}}
\def\e{{\textrm e}}
\def\del{\partial}
\def\half{{\frac12}}

\def\abs#1{{\left|{#1}\right|}}

\def\del{\partial}
\def\half{{\frac12}}

\def\abs#1{{\left|{#1}\right|}}

\def\del{\partial}
\def\dslash{\del\kern-0.55em\raise 0.14ex\hbox{/}}

\def\rough#1{\raise.3ex\hbox{$#1$\kern-.75em\lower1ex\hbox{$\sim$}}}

\newcommand{\PRD}[3]{{\it Phys. Rev.} {\bf D{#1}} (19{#3}) {#2}}

\newcommand{\PRDM}[3]{{\it Phys. Rev.} {\bf D{#1}} {#2} (20{#3})}

\newcommand{\PTP}[3]{{\it Prog. Theor. Phys.} {\bf {#1}} (19{#3}) {#2}}
\newcommand{\PTPM}[3]{{\it Prog. Theor. Phys.} {\bf {#1}} (20{#3}) {#2}}

\newcommand{\hepph}[1]{{\tt hep-ph/#1}}

\begin{document}
\begin{flushright}
{\small KOBE-COSMO-26-01}\\%
\end{flushright}
\begin{center}
{\LARGE\bf 
Nonanalytic Structure of Effective Potential at Finite Temperature \\ on Compactified Space
}
\vskip 1.4cm
Makoto Sakamoto$^{(a)}$
and Kazunori Takenaga$^{(b)}$
\footnote{E-mail: takenaga@kumamoto-hsu.ac.jp}
\\
\vskip 1.0cm
${}^{(a)}$ {\it Department of Physics, Kobe University, 
Rokkodai Nada, Kobe, 657-8501 Japan}
\\[0.2cm]
${}^{(b)}$ {\it Faculty of Health Science, Kumamoto
Health Science University, Izumi-machi, Kita-ku, Kumamoto 861-5598, Japan}
\\
\vskip 1.5cm
\begin{abstract}
We thoroughly investigate nonanalytic terms in the finite-temperature effective 
potential in one-loop approximation on 
a $D$-dimensional spacetime, $S_{\tau}\times R^{D-(p+1)}\times \prod_{i=1}^p S_i^1$, using a mode recombination
formula. Such nonanalytic terms cannot be expressed as positive powers of field-dependent mass squared. 
The formula provides a clear separation of the effective potential into a part that contains the nonanalytic terms
and a part that is purely analytic, and clarifies the origin of the nonanalytic terms. We obtain all the nonanalytic terms and 
show that only two types of nonanalytic terms arise from the modes with zero Matsubara frequency.
For a real scalar field with periodic boundary conditions, if the number of noncompacted spatial 
dimensions is odd (even), there are odd powers of $M$ ($\log M$ terms) but no $\log M$ 
terms (no odd powers of $M$). For fermions with general boundary conditions, we find that neither of the two 
types appears. These results clarify the nonanalytic structure of the finite-temperature effective 
potential on the spacetime with compactified spatial dimensions.
\end{abstract}
\end{center}
\vskip 1.0cm
\newpage
%
%
%
%
\section{Introduction}
%
%
Quantum field theory at finite temperature provides a fundamental framework for studying phase-transition phenomena 
in high-energy physics and has played a central role in advancing our understanding of the subject. In their 
pioneering analysis of the finite-temperature effective potential, Dolan and Jackiw\cite{dj} found that the 
scalar effective potential contains terms that cannot be expressed as positive powers of the field-dependent 
mass squared, namely, nonanalytic terms. The terms act as the source of the first-order phase transitions\cite{dine,quiros} and their 
magnitude directly controls the strength of the transition. They are, for instance, indispensable in electroweak 
baryogenesis scenario\cite{baryon} relying on the first-order electroweak phase transition. In exploring 
cosmological evolution scenarios\cite{trodden}, it is necessary to understand how symmetries 
are broken during the cooling of the Universe. In this respect, the study of nonanalytic contributions to the finite-temperature 
effective potential plays a crucial role. Nonanalytic terms are therefore of substantial physical significance.

Quantum field theories with compactified extra dimensions constitute one of the most promising approaches for exploring physics 
beyond the Standard Model. Gauge-Higgs unification provides a representative example\cite{hosotani}, where, once again, nonanalytic 
terms at finite-temperature effective potential are crucial for realizing the first-order phase transition\cite{marutake}. At the 
same time, theories with compact extra dimensions also serve as powerful theoretical tools for field-theoretic 
studies, providing a well-defined framework for such investigations. Dimensional reduction 
offers a prime example\cite{cava1,cava2} and moreover, these theories 
are known to exhibit rich and interesting phase structures\cite{host1,sakatake}.

In our previous two papers, referred to as I \cite{sakatake2022} and II \cite{sakatake2024}, we determined all the power-type 
nonanalytic terms in the one-loop finite-temperature effective potential 
of a scalar field obeying periodic boundary conditions on a $D$-dimensional 
spacetime, $S_{\tau}^1\times R^{D-(p+1)}\times \prod_{i=1}^pS_i^1$. In particular, we developed a new important formula 
called a mode recombination formula in the paper II and this formula enables a reformulation of the finite-temperature effective 
potential, thereby elucidating the origin of the nonanalytic terms. Although 
the origin of the nonanalytic terms is initially expressed using the modified 
Bessel functions of the second kind with multiple mode summations, we showed 
that, by employing the integral representations for the function together with the analytical extension of the mode summations, it can 
be recast as contour integrals in the complex plane. The subsequent evaluation of the relevant 
residues makes it possible to extract all the power-type nonanalytic terms in a straightforward manner.
The mode recombination formula clearly shows that the
emergence of nonanalytic terms is closely related to the existence of zero 
modes, making it a powerful tool for understanding the physical origin of such terms. As 
an immediate corollary, we also showed that fermion fields, whose 
zero modes are absent due to their statistics, exhibit no nonanalytic terms.

In this paper we derive a new remarkable expression of the finite-temperature effective potential 
in one-loop approximation by using extensively the mode recombination formula. The formula enables 
a clear separation of the effective potential into a part that contains nonanalytic terms and 
a part that remains purely analytic ones. We demonstrate that the effective potential possesses 
two physically important types of nonanalytic terms with respect to the field-dependent 
mass: power type and logarithmic one. These nonanalytic terms are obtained by the evaluation 
of the residues in a systematic and comprehensive manner. In the case of a real scalar 
field obeying the periodic boundary condition for the 
spatial $S_i^1~(i=1,\cdots, p)$ direction, we find that the nonanalytic structure of the 
finite-temperature effective potential turns out to be highly characteristic; the two types never appear simultaneously. 
%
%
%

We further show that, for fermion fields with arbitrary boundary conditions for the spatial $S_i^1$ direction, the 
application of the mode recombination formula renders the finite-temperature effective potential into a form in 
which nonanalytic terms can be analyzed more transparently. As a consequence, in addition to the absence of the power-type
nonanalytic terms as shown in the paper II, we find that the logarithmic-type nonanalytic term does not appear either
regardless of the values of $D$ and $p+1$. The result in the case of the fermion also holds for the case of 
the scalar that satisfies the antiperiodic boundary condition for at least one spatial $S_i^1$ direction.

The organization of this paper is as follows. Section $2$ provides the setup for analyzing nonanalytic 
terms in the finite-temperature effective potential and summarizes the method of the 
mode recombination formula developed in the paper II, followed by its application to the effective potential.
In section $3$ we determine all the nonanalytic terms in the effective potential of the scalar field with the 
periodic boundary conditions. Section $4$ presents the corresponding analysis for the fermion fields.
Section $5$ contains conclusions and discussions. We derive conditions under which evaluations of 
integrals by residues are applicable in Appendix A and demonstrate the evaluations in Appendixes B and C.
%
%
%
%
%
%
%
%
%
%
%
%
%
%
%
%
%
\section{Setup and Mode Recombination Formula}
%
%
Let us present the setup for the discussion in this paper and the mode recombination formula which
plays the important role to study the nonanalytic terms in the finite-temperature effective potential on the spacetime 
with compactified spatial dimensions. Here, we refer to the contents of those 
in sections $2$ and $3$ of the paper II. The familiar readers can directly go to the next section. 
\subsection{Setup}
\label{setup}
We first study the effective potential at finite temperature
in one-loop approximation on the $D$-dimensional spacetime, $S_{\tau}^1\times R^{D-(p+1)}\times \prod_{i=1}^p S_i^1$
for real scalar and fermion fields. We employ the Euclidean time formulation for quantum 
field theory at finite temperature. Then, the Euclidean time direction is compactified on $S_{\tau}^1$ whose coordinate 
and circumference are denoted by $\tau$ and $L_0$, respectively. The $L_0$ is the inverse 
temperature, $L_0=T^{-1}$. The spatial $p$ directions are compactified on the $p$ numbers 
of $S^1$.  Their coordinates are $y^i (i=1,\cdots, p)$ and 
$L_i$ is the circumference of each $S_i^1$. The $R^{D-(p+1)}$ is the 
flat Euclidean space and its coordinate is $x^k(k=1,\cdots, D-(p+1))$.

The Lagrangian is given by
\begin{align}
 {\cal L}&=\half (\del_N \phi)^2 -\frac{m_s^2}{2}\phi^2 -\frac{\lambda}{4!}\phi^4
 +\bar \psi(i\Gamma_N\del_N +m_f)\psi + g\phi\bar\psi\psi\,,
 \label{2.1}
 \end{align}
where the $N$ stands for $N=(\tau, k, i)$ and $\phi\,(\psi)$ is the scalar (fermion) field whose bulk mass is
$m_s\,(m_f)$. The $g$ is the Yukawa coupling.

We specify the boundary conditions for the $S_{\tau}^1$ and $S^1_i  (i=1,\cdots, p)$ directions. The boundary 
condition for the $S_{\tau}^1$ direction is defined by 
\begin{align}
\Phi(\tau+L_0, x^k, y^i)=\e^{2\pi i\eta_0}\Phi(\tau, x^k, y^i)
 \label{2.2}
\end{align}
for a given field $\Phi(\tau, x^k, y^i)$. The parameter $\eta_0$ is definitely determined by 
quantum statistics to be $0$ (periodic) for the scalar field or to be
$\half$ (antiperiodic) for the fermion field. On the other hand, the boundary condition for the spatial 
$S_i^1$ direction is imposed by
\begin{align}
\Phi(\tau, x^k, y^i+L_i)&=\e^{2\pi i\eta_i}\Phi(\tau, x^k, y^i),
 \label{2.3}
\end{align}
where the parameter $\eta_i$ can take $0$ or $\half$ for the real scalar field and can be arbitrary for the fermion field.

We employ the standard prescription to calculate the effective potential at finite temperature in 
one-loop approximation. By taking up the quadratic terms in the shifted Lagrangian around the constant field $\varphi$
for the scalar field $\phi$ in Eq.(\ref{2.1}), one needs to evaluate 
\begin{align}
V_{\rm eff}
&=(-1)^{f}{\cal N}~\half\left( \prod_{i=0}^{p}\frac{1}{L_i}\sum_{n_i=-\infty}^{\infty}\right)
\int\frac{d^{D-(p+1)}p_E}{(2\pi)^{D-(p+1)}}
   \notag\\
&\hspace{5mm}\times
\log\Bigl[p_E^2 +\left(\frac{2\pi}{L_0}\right)^2(n_0 +\eta_0)^2
+\sum_{i=1}^p\left(\frac{2\pi}{L_i}\right)^2(n_i +\eta_i)^2
+M^2(\varphi)\Bigr]
 \label{2.4}
\end{align}
in order to obtain the effective potential on $S_{\tau}^1\times R^{D-(p+1)}\times \prod_{i=1}^{p}S^1_i$
in one-loop approximation. Here, the $M(\varphi)$ is the field-dependent mass 
of the scalar (fermion) field
\begin{align}
M^2(\varphi)=m_s^2+\frac{\lambda}{2}\varphi^2 \quad \Bigl(M(\varphi)=m_f+g\varphi\Bigr).
 \label{2.5}
\end{align}
Hereafter, we denote $M(\varphi)$ by $M$ for simplicity. The $p_E$ is the $D-(p+1)$-dimensional 
Euclidean momentum. The $f$ is the fermion number that is $0~(1)$ for the boson (fermion) and 
the ${\cal N}$ is the on-shell degrees of freedom. The $n_{0}$ stands for the Matsubara mode arising 
from the $S_{\tau}^1$ and the Kaluza-Klein mode $n_i~(i=1,\cdots, p)$ comes from each $S_i^1~(i=1,\cdots p)$. The 
parameter $\eta_i\, (i=0,1,\cdots, p)$ is given in Eqs.(\ref{2.2}) and (\ref{2.3}).

We make use of the zeta-function regularization in order to evaluate Eq.(\ref{2.4}). We can write 
\begin{align}
V_{\rm eff}=(-1)^f{\cal N}\half\left(-\frac{d}{ds}I(s)\right)\Bigg|_{s\rightarrow 0},
 \label{2.6}
\end{align}
where 
\begin{align}
I(s)&\equiv \left( \prod_{i=0}^{p}\frac{1}{L_i}\sum_{n_i=-\infty}^{\infty}\right)
\int\frac{d^{D-(p+1)}p_E}{(2\pi)^{D-(p+1)}}\notag\\
&\hspace{5mm}
\times 
\Bigl[p_E^2 +\left(\frac{2\pi}{L_0}\right)^2(n_0 +\eta_0)^2
+\sum_{i=1}^p\left(\frac{2\pi}{L_i}\right)^2(n_i +\eta_i)^2
+M^2\Bigr]^{-s}.
 \label{2.7}
\end{align}
We first use the formula
\begin{align}
A^{-s}=\frac{1}{\Gamma(s)}\int_0^{\infty}dt~t^{s-1}\e^{-At}
 \label{2.8}
\end{align} 
and perform the Gaussian $p_E$ integration. Then, we employ the Poisson summation
\begin{align}
\sum_{n_j=-\infty}^{\infty}\e^{-(\frac{2\pi}{L_j})^2(n_j+\eta_j)^2t}=
\sum_{m_j=-\infty}^{\infty}\frac{L_j}{2\pi}\left(\frac{\pi}{t}\right)^{\half}
\e^{-\frac{(m_jL_j)^2}{4t}+2\pi im_j\eta_j}
 \label{2.9}
\end{align}
to obtain
\begin{align}
V_{\rm eff}&=(-1)^{f+1}{\cal N}\frac{\pi^{\frac{D}{2}}}{2(2\pi)^D}
\sum_{m_0=-\infty}^{\infty}\cdots \sum_{m_p=-\infty}^{\infty}
\notag \\
&\hspace{5mm}\times
  \int_0^{\infty}dt~t^{-\frac{D}{2}-1}\e^{-\frac{1}{4t}[(m_0L_0)^2+\cdots+(m_pL_p)^2] -M^2t +2\pi i(m_0\eta_0+\cdots +m_p\eta_p)}.
   \label{2.10}
\end{align}
We call $m_j~(j=0, 1,\cdots, p)$ in Eq.(\ref{2.9}) the winding modes, while $n_0$ and $n_i~(i=1,2,\cdots, p)$ the 
Matsubara and the Kaluza-Klein modes, respectively.

It is convenient to separate each sum over $m_j$ in Eq.(\ref{2.10}) into the zero winding
mode $(m_j=0)$ and the nonzero winding ones $(m_j\neq 0)$ and to express Eq.(\ref{2.10}) into the  form
\begin{align}
V_{\text{eff}} = \sum_{n=0}^{p+1} F^{(n)D}(M)=
\sum_{n=0}^{p+1}~\sum_{0\leq i_1<i_2<\cdots < i_n\leq p}F^{(n)D}_{L_{i_1},L_{i_2},\cdots, L_{i_n}}(M),
 \label{2.11}
\end{align}
where
\begin{align}
F^{(n)D}_{L_{i_1},L_{i_2},\cdots, L_{i_n}}(M)
 &= (-1)^{f+1}{\cal N}\frac{\pi^{\frac{D}{2}}}{2(2\pi)^D}
     \sum_{m_{i_1}=-\infty}^{\infty}{}^{\hspace{-3mm}\prime}\cdots 
     \sum_{m_{i_n}=-\infty}^{\infty}{}^{\hspace{-3mm}\prime}
     \notag \\
 &\hspace{5mm}\times
    \int_0^{\infty}dt~t^{-\frac{D}{2}-1}\e^{-\frac{1}{4t}[(m_{i_1}L_{i_1})^2+
     \cdots+(m_{i_n}L_{i_n})^2]-M^2t +2\pi i(m_{i_1}\eta_{i_1}+\cdots +m_{i_n}\eta_{i_n})}.
      \label{2.12}
\end{align}
The prime of the summation in $\sum_{m_j=-\infty}^{\,\prime\, \infty}$ means that the zero winding mode ($m_j=0$) is 
removed. By using the formula 
\begin{align}
\int_0^{\infty}dt~t^{-\nu-1}\e^{-At-\frac{B}{t}}=2\left(\frac{A}{B}\right)^{\frac{\nu}{2}}K_{\nu}(2\sqrt{AB}),
  \label{2.13}
\end{align}
where the $K_{\nu}(z)$ is the modified Bessel function of the second kind, Eq.(\ref{2.12}) for $n\geq 1$ becomes
\begin{align}
F_{L_{i_1},L_{i_2},\cdots, L_{i_{n}}}^{(n)D}(M)
 &= (-1)^{f+1}{\cal N}\frac{2^n}{(2\pi)^{\frac{D}{2}}}
    \sum_{m_{i_{1}}=1}^{\infty} \cdots \sum_{m_{i_{n}}=1}^{\infty}
    \left(\frac{M^2}{(m_{i_1}L_{i_1})^2+\cdots +(m_{i_n}L_{i_n})^2}\right)^{\frac{D}{4}}
    \notag\\
&\hspace{5mm}\times 
    K_{\frac{D}{2}}\left(\sqrt{{M^2}\{(m_{i_1}L_{i_1})^2+\cdots +(m_{i_n}L_{i_n})^2\}}\right)\notag\\
    &\hspace{15mm}\times
    \cos(2\pi m_{i_1}\eta_{i_1})\cdots \cos(2\pi m_{i_n}\eta_{i_n}).
      \label{2.14}
\end{align}

$F^{(0)D}$ in Eq.(\ref{2.11}) is the contribution from
all the zero winding modes $m_{0} = m_{1} = \cdots = m_{p} = 0$ in Eq.(\ref{2.10}) and is calculated to be
\begin{align}
F^{(0)D}(M)
 = (-1)^{f+1}{\cal N}\frac{\pi^{\frac{D}{2}}}{2(2\pi)^D}
    \int_0^{\infty}dt~t^{-\frac{D}{2}-1}\e^{-M^2t}
 = (-1)^{f+1}{\cal N}\frac{\pi^{\frac{D}{2}}}{2(2\pi)^D}
   \Gamma(-\tfrac{D}{2})(M^2)^{\frac{D}{2}},
     \label{2.15}
\end{align}
where we have used Eq.(\ref{2.8}).  For $D=$ even, if one makes use of the 
dimensional regularization and $\overline{MS}$ renormalization, for example, then, we obtain 
\begin{align}
F^{(0)D}(M)=(-1)^{f+1}{\cal N}
\frac{(-1)^{\frac{D}{2}}}
{2^{D+1}\pi^{\frac{D}{2}}(\frac{D}{2})!}
M^D
\Bigl\{-\log\frac{M^2}{\mu^2}+\sum_{r=1}^{\frac{D}{2}}\frac{1}{r}
\Bigr\},
 \label{2.16}
\end{align}
where the scale $\mu$ is a sequence of possible renormalization conditions. On the other hand, for $D=$ odd, with the help
of the formula 
\begin{align}
\Gamma\left(-\frac{D}{2}\right)\Bigg|_{D={\text {odd}}}=\frac{(-1)^{\frac{D+1}{2}}2^{\frac{D+1}{2}}}{D!!}\sqrt{\pi},
\label{2.17}
\end{align}
we have
\begin{align}
F^{(0)D}(M)=(-1)^{f+1}{\cal N}\frac{(-1)^{\frac{D+1}{2}}}{2^{\frac{D+1}{2}}\pi^{\frac{D-1}{2}}D!!}M^D.
\label{2.18}
\end{align}
For $D$=even (odd), $F^{(0)D}(M)$ has the logarithmic(power)-type nonanalytic 
terms, as shown in Eqs.(\ref{2.16}) and (\ref{2.18}).
%
\subsection{Mode Recombination Formula}
\label{mrf}
%
The mode recombination formula developed in the paper II plays the important role 
to study the nonanalytic terms. We give a brief review of the formula here for the reader's convenience.

Let us first recast Eq.(\ref{2.11}) into 
\begin{align}
V_{\text{eff}} &= F^{(0)D}(M)+
\sum_{n=1}^{p+1}~\sum_{0\leq i_1<i_2<\cdots < i_n\leq p}F^{(n)D}_{L_{i_1},L_{i_2},\cdots, L_{i_n}}(M)\notag\\
&=
F^{(0)D}(M)\notag\\
&\hspace{5mm}
+
\Big\{
F_{L_0}^{(1)D}(M)+
\sum_{n=1}^{p}~\sum_{1\leq i_1<i_2<\cdots < i_n\leq p}
\Bigl(F_{L_{i_1},\cdots, L_{i_n}}^{(n)D}(M)+F_{L_0, L_{i_1},\cdots, L_{i_n}}^{(n+1)D}(M)\Bigr)
\Big\}.
\label{2.19}
\end{align}
The terms in the curly brackets\footnote{$F_{L_0}^{(1)D}(M)$ corresponds to the finite-temperature effective potential obtained by 
Dolan and Jackiw\cite{dj}.} are 
obtained by focusing on the scale $L_0$ in $L_{i_j} (i_j=0,\cdots,p)$ and separating the term 
associated with $L_0$ from the one without $L_0$ on the right-hand side of Eq.(\ref{2.11}). We have proved in the paper II
\begin{align}
F_{L_{i_1},\cdots, L_{i_n}}^{(n)D}(M)+F_{L_0, L_{i_1},\cdots, L_{i_n}}^{(n+1)D}(M)=
\frac{1}{L_0}\sum_{n_0=-\infty}^{\infty}F_{L_{i_1},\cdots, L_{i_n}}^{(n)D-1}(M_{(0)}),
  \label{2.20}
\end{align}
where
\begin{align}
M_{(0)}^2\equiv M^2+\left(\frac{2\pi}{L_0}\right)^2(n_0+\eta_0)^2.
  \label{2.21}
\end{align}
We call Eq.(\ref{2.20}) the mode recombination formula and the formula holds irrespective of whether
fields are fermions or scalars and also of boundary conditions for the spatial $S_i^1~(i=1,\cdots, p)$ direction.
It should be noted that, as shown in the paper II, the Matsubara mode $n_0$ is recovered 
instead of the winding mode $m_0$ by using the Poisson summation (\ref{2.9}) inversely.

By using Eq.(\ref{2.20}), Eq.(\ref{2.19}) becomes
\begin {align}
V_{\text{eff}} =
F^{(0)D}(M)+\Big\{
F_{L_0}^{(1)D}(M)+
\sum_{n=1}^{p}~\sum_{1\leq i_1<i_2<\cdots < i_n\leq p}
\frac{1}{L_0}\sum_{n_0=-\infty}^{\infty}F_{L_{i_1},\cdots, L_{i_n}}^{(n)D-1}(M_{(0)})
\Big\}.
\label{2.22}
\end{align}
We can repeat the same procedure by focusing next on the scales $L_1, L_2,\cdots, L_{p-1}$ sequentially for the term with 
the multiple mode summations like the third term in Eq.(\ref{2.22}) and 
we have finally arrived at the expression (3.11) in the paper II for the effective potential $V_{\text{eff}}$. 
Moreover, we have also shown in the paper II that the first and second terms in Eq.(\ref{2.22}) can be recast into
\begin{align}
F^{(0)D}(M)+F_{L_0}^{(1)D}(M)=\frac{1}{L_0}\sum_{n_0=-\infty}^{\infty}F^{(0)D-1}(M_{(0)}),
\label{2.23}
\end{align}
which formally follows from the formula (\ref{2.20}) with $n=0$. Then, one has
\begin{align}
V_{\text{eff}} =\frac{1}{L_0}\sum_{n_0=-\infty}^{\infty}F^{(0)D-1}(M_{(0)})
+\sum_{n=1}^{p}~\sum_{1\leq i_1<i_2<\cdots < i_n\leq p}
\frac{1}{L_0}\sum_{n_0=-\infty}^{\infty}F_{L_{i_1},\cdots, L_{i_n}}^{(n)D-1}(M_{(0)}).
\label{2.24}
\end{align}
Eqs.(\ref{2.22}) and (\ref{2.24}) have already been obtained in the paper II and we rewrite them into 
more convenient forms in the next section, making the subsequent discussion more transparent.

Eq.(\ref{2.14}) contains the modified Bessel function of the second kind involving  
the multiple mode summations. The integral representation for the function 
in the complex plane is the powerful tool to study the nonanalytic terms in the effective potential as shown in
the papers I and II. We also follow the same strategy in the present paper and introduce 
the integral representation for the modified Bessel function of the second kind in the complex plane\cite{bromwich, davis}
\begin{align}
K_{\nu}(x)=\frac{1}{4\pi i}\int_{c-i\infty}^{c+i\infty}dt~\Gamma(t)\Gamma(t-\nu)\left(\frac{x}{2}\right)^{-2t+\nu}.
\label{2.25}
\end{align}
Here, the constant $c$ should be understood to be a point located on the real axis which is greater than all 
the poles of the gamma functions in the integrand.

Applying the formula (\ref{2.25}) to Eq.(\ref{2.14}) together with Eq.(\ref{2.21}), we obtain the 
integral form in the complex plane for the third term of Eq.(\ref{2.22}) 
\begin{align}
&\frac{1}{L_0}\sum_{n_0=-\infty}^{\infty}F_{L_{i_1},\cdots, L_{i_n}}^{(n)D-1}(M_{(0)})\notag\\
&\hspace{5mm}
=(-1)^{f+1}{\cal N}\frac{2^n\pi^{\frac{D-1}{2}}}{L_0^D}
\frac{1}{4\pi i}
\int_{c-i\infty}^{c+i\infty}dt~\Gamma(t-\tfrac{D-1}{2})\sum_{n_0=-\infty}^{\infty}
\Biggl\{
(n_0+\eta_0)^2 +\left(\frac{ML_0}{2\pi}\right)^2
\Biggr\}^{-t+\frac{D-1}{2}}\notag\\
&\hspace{15mm}\times
\left(\frac{\pi}{L_0}\right)^{-2t}S^{(n)}(t; L_{i_1},\cdots, L_{i_n}),
\label{2.26}
\end{align}
where we have defined 
\begin{align}
S^{(n)}(t; L_{i_1},\cdots, L_{i_n})&\equiv \Gamma(t)\sum_{m_{i_1}=1}^{\infty}\cdots\sum_{m_{i_n}=1}^{\infty}
\left\{
(m_{i_1}L_{i_1})^2+\cdots +(m_{i_n}L_{i_n})^2
\right\}^{-t}\notag\\
&\hspace{5mm}\times \cos(2\pi m_{i_1}\eta_{i_1})\cdots \cos(2\pi m_{i_n}\eta_{i_n}).
\label{2.27}
\end{align} 
The right-hand side of Eq.(\ref{2.23}) is recast by using Eq.(\ref{2.15}) with Eq.(\ref{2.21}) into
\begin{align}
\frac{1}{L_0}\sum_{n_0=-\infty}^{\infty}F^{(0)D-1}(M_{(0)})&=\frac{1}{L_0}(-1)^{f+1}{\cal N}
\frac{\pi^{\frac{D-1}{2}}}{2(2\pi)^{D-1}}\Gamma(-\tfrac{D-1}{2})\sum_{n_0=-\infty}^{\infty}(M_{(0)}^2)^{\frac{D-1}{2}}\notag\\
&=(-1)^{f+1}{\cal N}\frac{\pi^{\frac{D-1}{2}}}{2L_0^D}\Gamma(-\tfrac{D-1}{2})\sum_{n_0=-\infty}^{\infty}
\Bigg\{(n_0+\eta_0)^2+\left(\frac{ML_0}{2\pi}\right)^2\Biggr\}^{\frac{D-1}{2}}.
\label{2.28}
\end{align}
The mode sum over $n_0$ in Eqs.(\ref{2.26}) and (\ref{2.28}) has the analytical 
extension\cite{elizalde}. Depending on the value of $\eta_0$ that specifies the boundary condition for the
$S_{\tau}^1$ direction, the form of the analytical extension is different and it is preferable to treat 
the fermion and the scalar separately.

Eqs.(\ref{2.22}) and (\ref{2.26}) are the starting point in the present paper to
investigate the nonanalytic structure of the effective potential for the case of the scalar field with 
the periodic boundary condition for the spatial $S_i^1$ direction. 
On the other hand, Eqs.(\ref{2.24}), (\ref{2.26}) and (\ref{2.28}) will be 
used for studying the nonanalytic terms in the effective potential for the case 
of the fermion field with arbitrary boundary condition for the spatial $S_i^1$ direction.
%
%
%
%
\section{Nonanalytic terms for scalar field with periodic \\ boundary condition} 
In this section we analyze the nonanalytic structure of the effective potential $V_{\text{eff}}$ in the case of a real scalar 
field $(f=0, \eta_0=0, {\cal N}=1)$ with the periodic boundary condition $(\eta_1=\cdots = \eta_p=0)$ for the
spatial $S^1_i~(i=1,\cdots, p)$ direction. To this end, we further recast Eq.(\ref{2.22}) into a more convenient form.
\subsection{New form of $V_{\text{eff}}$}
We separately write the zero mode $(n_0=0)$ part and the nonzero mode $(n_0\neq 0)$ one 
in Eq.(\ref{2.26}) to yield 
\begin{align}
\frac{1}{L_0}\sum_{n_0=-\infty}^{\infty}F_{L_{i_1},\cdots, L_{i_n}}^{(n)D-1}(M_{(0)})
&=\frac{1}{L_0}F_{L_{i_1},\cdots, L_{i_n}}^{(n)D-1}(M)
-\frac{2^n\pi^{\frac{D-1}{2}}}{L_0^D}
\frac{1}{2\pi i}
\int_{c-i\infty}^{c+i\infty}dt~\Gamma(t-\tfrac{D-1}{2})\notag\\
&\hspace{5mm}\times 
\sum_{n_0=1}^{\infty}
\Biggl\{
n_0^2 +\left(\frac{ML_0}{2\pi}\right)^2
\Biggr\}^{-t+\frac{D-1}{2}}
\left(\frac{\pi}{L_0}\right)^{-2t}{\tilde S}^{(n)}(t; L_{i_1},\cdots, L_{i_n}),
\label{3.1}
\end{align}
where we have defined 
\begin{align}
{\tilde S}^{(n)}(t; L_{i_1},\cdots, L_{i_n})=\Gamma(t)\sum_{m_{i_1}=1}^{\infty}\cdots\sum_{m_{i_n}=1}^{\infty}
\Bigl\{
(m_{i_1}L_{i_1})^2+\cdots +(m_{I_n}L_{i_n})^2
\Bigr\}^{-t}.
\label{3.2}
\end{align}

The mode sum over $n_0$ in Eq.(\ref{3.1}) has the analytical extension\cite{elizalde} and it is given by 
\begin{align}
\Gamma(t)\sum_{n_0=1}^{\infty}\{n_0^2+z^2\}^{-t}
&=
-\half  \frac{\Gamma(t)}{z^{2t}} 
+ \frac{\sqrt{\pi}}{2}\frac{\Gamma(t-\half)}{z^{2(t-\half)}}\nonumber\\
&\hspace{5mm}
+\frac{\pi^{2t}}{\sqrt{\pi}}
\frac{1}{2\pi i}\int_{c_{1} - i\infty}^{c_{1} +i\infty}d{t_{1}}~\Gamma(t_{1} -t +\tfrac{1}{2})
\zeta(2t_{1}-2t +1)\Gamma(t_{1}) (\pi z)^{-2t_{1}}.
\label{3.3}
\end{align}
After changing the variable $\bar t=t-\tfrac{D-1}{2}$ in Eq.(\ref{3.1}) and then applying Eq.(\ref{3.3}) to the second term in Eq.(\ref{3.1}), we obtain 
\begin{align}
&-\frac{2^n}{L_0\pi^{\frac{D-1}{2}}}\frac{1}{2\pi i}\int_{c-\frac{D-1}{2}-i\infty}^{c-\frac{D-1}{2}+i\infty}dt~\Biggl\{
-\frac{1}{2}\Gamma(t)\left(\frac{ML_0}{2\pi}\right)^{-2t}+\frac{\sqrt{\pi}}{2}\Gamma(t-\tfrac{1}{2})\left(\frac{ML_0}{2\pi}\right)^{-2(t-\frac{1}{2})}
\notag\\
&\hspace{5mm}+\frac{\pi^{2t}}{\sqrt{\pi}}\frac{1}{2\pi i}\int_{c_1-i\infty}^{c_1+i\infty}dt_1~\Gamma(t_1-t+\tfrac{1}{2})\zeta(2t_1-2t+1)\Gamma(t_1)
\left(\frac{ML_0}{2}\right)^{-2t_1}
\Biggr\}\notag\\
&\hspace{10mm}\times \left(\frac{\pi}{L_0}\right)^{-2t}{\tilde S}^{(n)}(t+\tfrac{D-1}{2}; L_{i_1},\cdots, L_{i_n}),
\label{3.4}
\end{align}
where we have denoted $\bar t$ as $t$ again.

We deform the $t_1$ integration path in Eq.(\ref{3.4}) to encircle 
all the poles of the integrand and evaluate the $t_1$ integral by the residue theorem. 
For this to be possible, the condition
\begin{align}
0 < \frac{ML_0}{2\pi} < 1
\label{3.5}
\end{align}
must be satisfied. This follows from the requirement that the contribution from the large left 
semicircle attached to the original $t_1$ integral must vanish for the residue theorem to apply.
The condition (\ref{3.5}) will be derived in Appendix A.

Among the poles $t_1=t-\half -\ell~(\ell=0, 1, \cdots)$ of $\Gamma(t_1-t+\tfrac{1}{2})$, only the pole
$t_1=t-\tfrac{1}{2}$ is relevant because of $\zeta(-2\ell)=0~(\ell=1,2,\cdots)$, which always follows from the 
combination $\Gamma(t_1-t+\tfrac{1}{2})\zeta(2t_1-2t+1)$ and is frequently used throughout 
our discussions. The residue at the pole $t_1=t-\tfrac{1}{2}$ with $\zeta(0)=-1/2$ cancels the second 
term in Eq.(3.4) and that at the pole $t_1=t$ of $\zeta(2t_1-2t+1)$ with $\Gamma(\tfrac{1}{2})=\sqrt{\pi}$ does the 
first term in Eq.(3.4). Thus, we are left with the contribution from the pole $t_1=-\bar n~(\bar n=0,1,\cdots)$ 
of $\Gamma(t_1)$ and Eq.(\ref{3.1}) becomes\footnote{We exchange the order of the sum over $\bar n$ and
the $t$ integral in Eq.(\ref{3.6}). The condition (\ref{3.5}) ensures the convergence of the sum, and one can show that
the $t$ integral along the imaginary axis is well defined by using the discussion in the Appendix of the paper I
together with the one in Appendix A of this paper. We also apply the exchange in subsequent equations, although
we do not state it explicitly each time. Since the technical details of the exchange lie outside the scope of this 
paper, we do not pursue them further.}
\begin{align}
&\frac{1}{L_0}\sum_{n_0=-\infty}^{\infty}F_{L_{i_1},\cdots, L_{i_n}}^{(n)D-1}(M_{(0)})\notag\\
&=\frac{1}{L_0}F_{L_{i_1},\cdots, L_{i_n}}^{(n)D-1}(M)\notag\\
&\hspace{5mm}
-\frac{2^n}{\pi^{\frac{D}{2}}L_0}
\sum_{\bar n=0}^{\infty}\frac{(-1)^{\bar n}}{\bar n!}\left(\frac{ML_0}{2}\right)^{2\bar n}
\frac{1}{2\pi i}\int_{c-\frac{D-1}{2} -i\infty}^{c-\frac{D-1}{2}+i\infty}dt~
\Gamma( -t-\bar n+\tfrac{1}{2})
\zeta( -2t-2\bar n +1)
\notag\\
&\hspace{10mm}\times \left(\frac{1}{L_0}\right)^{-2t}{\tilde S}^{(n)}(t+\tfrac{D-1}{2}; L_{i_1},\cdots, L_{i_n}).
\label{3.6}
\end{align}

There may appear the poles in the multiple mode summations, ${\tilde S}^{(n)}(t+\tfrac{D-1}{2}; L_{i_1},\cdots, L_{i_n})$ that
contribute to the above $t$ integral. It does not, however, change the $(M^2)^{\bar n}$ dependence
in Eq.(\ref{3.6}) because of the nonexistence of $M$ in ${\tilde S}^{(n)}(t+\tfrac{D-1}{2}; L_{i_1},\cdots, L_{i_n})$.
Hence, the second term in Eq.(\ref{3.6}), which arises from the nonzero modes $(n_0\neq 0)$, is 
entirely analytic with respect to $M$ and is expressed in the positive powers of $M^2$. 
This means that any nonanalytic term, if present, arises from the first term in Eq.(\ref{3.6}) that 
comes from the contribution of the zero mode $(n_0=0)$. We recognize that the zero mode is crucial for the 
existence of the nonanalytic term. The effective potential (\ref{2.22}) with Eq.(\ref{3.6}) becomes
\begin {align}
V_{\text{eff}} &=
F^{(0)D}(M)+F_{L_0}^{(1)D}(M)+\sum_{n=1}^{p}~\sum_{1\leq i_1<i_2<\cdots < i_n\leq p}
\frac{1}{L_0}F_{L_{i_1},\cdots, L_{i_n}}^{(n)D-1}(M)\notag\\
&
-\sum_{n=1}^{p}~\sum_{1\leq i_1<i_2<\cdots < i_n\leq p}
\frac{2^n}{\pi^{\frac{D}{2}}L_0}\sum_{\bar n=0}^{\infty}\frac{(-1)^{\bar n}}{\bar n!}\left(\frac{ML_0}{2}\right)^{2\bar n}
\frac{1}{2\pi i}\int_{c-\frac{D-1}{2} -i\infty}^{c-\frac{D-1}{2}+i\infty}dt~
\notag\\
&\hspace{5mm}\times 
\Gamma(-t-\bar n +\tfrac{1}{2})\zeta(-2t-2\bar n  +1)
\left(\frac{1}{L_0}\right)^{-2t}{\tilde S}^{(n)}(t+\tfrac{D-1}{2}; L_{i_1},\cdots, L_{i_n}).
\label{3.7}
\end{align}

We repeat the same discussion as the subsection \ref{mrf} focusing next on the scale $L_1$ 
for the third term in Eq.(\ref{3.7}) and apply the analytical extension (\ref{3.3}) to the mode sum over $n_1$ associated with the 
scale $L_1$. Then, we evaluate the $t_1$-integral like Eq.(\ref{3.4}) by the residue theorem to yield 
\begin{align}
&\sum_{n=1}^p\sum_{1\leq i_1<i_2<\cdots < i_n\leq p} F_{L_{i_1},\cdots, L_{i_n}}^{(n)D-1}(M)\notag\\
&=F_{L_1}^{(1)D-1}(M)+\sum_{n=1}^{p-1}\sum_{2\leq i_1<i_2<\cdots < i_n\leq p}\frac{1}{L_1} F_{L_{i_1},\cdots, L_{i_n}}^{(n)D-2}(M)
\notag\\
&\hspace{5mm}
-\sum_{n=1}^{p-1}\sum_{2\leq i_1<i_2<\cdots < i_n\leq p}
\frac{2^n}{\pi^{\frac{D-1}{2}}L_1}
\sum_{\bar n=0}^{\infty}\frac{(-1)^{\bar n}}{\bar n !} \left(\frac{ML_1}{2}\right)^{2\bar n}
\frac{1}{2\pi i}\int_{c-\frac{D-2}{2}-i\infty}^{c-\frac{D-2}{2}+i\infty}dt~
\notag\\
&\hspace{15mm}\times 
\Gamma(-t-\bar n +\tfrac{1}{2})\zeta(-2t-2\bar n+1)
\left(\frac{1}{L_1}\right)^{-2t}
\tilde S(t+\tfrac{D-2}{2}; L_{i_1},\cdots, L_{i_n}).
\label{3.8}
\end{align}
One needs the condition $0<ML_1/(2\pi) < 1$ for the $t_1$ integration to be evaluated by the residue 
theorem, which will be discussed in Appendix A. 
Let us note that in Eq.(\ref{3.8}) the second term is the contribution from 
the zero mode $(n_1=0)$, while the third one comes from the nonzero modes $(n_1\neq 0)$.
We observe again that the contribution from the nonzero modes depends on the positive powers of $M^2$, so that
it is analytical, and the nonanalytic terms potentially originate from the first and the second terms. This 
procedure can be continued in order up to $L_{p-1}$ by applying the mode recombination formula 
successively $p$ times to the term with the multiple mode summations 
like $F_{L_{i_1},\cdots, L_{i_n}}^{(n)D-2}(M)$ in Eq.(\ref{3.8}).

Then, we finally derive a new remarkable expression for the effective potential
\begin {align}
V_{\text{eff}} &=F^{(0)D}(M)+\sum_{k=0}^p\left(\frac{1}{\prod_{i=0}^{k-1}L_i}\right)F_{L_k}^{(1)D-k}(M)
\notag\\
&
-\sum_{k=1}^p\sum_{n=1}^{p+1-k}~\sum_{k\leq i_1<i_2<\cdots < i_n\leq p}
\frac{2^n}{\pi^{\frac{D-(k-1)}{2}}}\frac{1}{\prod_{j=0}^{k-1}L_j}
\sum_{\bar n=0}^{\infty}\frac{(-1)^{\bar n}}{\bar n!}\left(\frac{ML_{k-1}}{2}\right)^{2\bar n}
\frac{1}{2\pi i}\int_{c-\frac{D-k}{2} -i\infty}^{c-\frac{D-k}{2}+i\infty}dt~
\notag\\
&\hspace{5mm}\times 
\Gamma(-t-\bar n +\tfrac{1}{2})\zeta(-2t-2\bar n  +1)
\left(\frac{1}{L_{k-1}}\right)^{-2t}{\tilde S}^{(n)}(t+\tfrac{D-k}{2}; L_{i_1},\cdots, L_{i_n}).
\label{3.9}
\end{align}
It should be understood that $\prod_{i=0}^{k-1}L_i=1$ for $k=0$ in the second term of Eq.(\ref{3.9}).
We observe from Eq.(\ref{3.9}) that the mode recombination formula 
enables the clear separation between the terms that contain the nonanalytic terms
and entirely analytic ones. Hence, we conclude that the nonanalytic terms in $V_{\text{eff}}$ are given by
\begin{align}
V_{\text{eff}}\big|_{\text{n.a.}}=F^{(0)D}(M)\Big|_{\text{n.a.}}+\sum_{k=0}^p
\left(\frac{1}{\prod_{i=0}^{k-1}L_i}\right)F_{L_k}^{(1)D-k}(M)\Big|_{\text{n.a.}},
\label{3.10}
\end{align}
where the abbreviation denoted by ``n.a." means {\it nonanalytic terms}. 
Let us remind the reader again that the second term in Eq.(\ref{3.10}) is the contributions from 
the zero modes $(n_0=n_1=\cdots=n_p=0)$.

We repeatedly evaluate the $t_1$ integral of the type appearing in Eq.(\ref{3.4}) in order to obtain Eq.(\ref{3.9}).
For this to be possible, one needs the condition
\begin{align}
0< \frac{ML_i}{2\pi}< 1~~(i=0, 1, \cdots, p).
\label{3.11}
\end{align}
From the above argument alone, we obtain only the
conditions up to $i=p-1$, but as we will see below and in Appendix A, Eq.(\ref{3.11}) also specifies the conditions under 
which the residue theorem for computing the nonanalytic terms in Eq.(\ref{3.10}) can be applied.
\subsection{Nonanalytic terms}
The first term in Eq.(\ref{3.10}) is already calculated to be given by Eqs.(\ref{2.16}) and (\ref{2.18}). 
We analyze what types of nonanalytic terms appear in the second term of Eq.(\ref{3.10}). 
From Eqs.(\ref{2.14}) and (\ref{2.25}), $F_{L_k}^{(1)D-k}(M)$ can be recast as
\begin{align}
F_{L_k}^{(1)D-k}(M)&=\frac{-2}{(2\pi)^{\frac{D-k}{2}}}\sum_{m_k=1}^{\infty}\left(\frac{M^2}{(m_kL_k)^2}\right)^{\frac{D-k}{4}}
K_{\frac{D-k}{2}}(m_kML_k)\notag\\
&=\frac{-2}{(2\pi)^{\frac{D-k}{2}}}\left(\frac{M^2}{2}\right)^{\frac{D-k}{2}}
\frac{1}{4\pi i} \int_{c-i\infty}^{c+i\infty}dt~\Gamma(t-\tfrac{D-k}{2})\Gamma(t)\zeta(2t)\left(\frac{ML_k}{2}\right)^{-2t},
\label{3.12}
\end{align}
where we have used the definition of the zeta function $\zeta(z)=\sum_{m=1}^{\infty}m^{-z}$. By 
deforming the contour in the complex $t$ plane so as to enclose all the poles of the integrand, we evaluate the 
$t$ integral using the residue theorem. For this to be possible, as discussed in Appendix A, one needs the 
condition, $0< ML_k/(2\pi) < 1~(k=0,\cdots, p)$, which is the same as Eq.(\ref{3.11}).

In the paper II we are only interested in the power-type nonanalytic terms with respect 
to $M$, that is, the odd power of $M$ and calculate the integral by the 
residue theorem by focusing only on the single pole of the integrand. We did not studied other nonanalytic terms in the paper. In 
the present paper we thoroughly investigate other type of nonanalytic 
terms, so that the higher-order pole of the integrand must be taken into account. In fact, as we will see below, the double pole 
generates the logarithmic-type nonanalytic term, $\log M$.

Computing the $t$ integral by the residue theorem in Eq.(\ref{3.12}) is straightforward 
and the details of the calculation are given in Appendix B. We present the results here. 
For $(D, k)=({\text{even}}, {\text{even}}), ({\text{odd}}, {\text{odd}})$, we obtain 
\begin{align}
F_{L_k}^{(1)D-k}(M)&=
\frac{(-1)(-1)^{\frac{D-k}{2}}}{2^{\frac{D-k}{2}}\pi^{\frac{D-k-2}{2}}(D-k-1)!!}\frac{M^{D-k-1}}{L_k}\notag\\
&\hspace{5mm}+
\frac{(-1)(-1)^{\frac{D-k}{2}}}{2^{D-k}\pi^{\frac{D-k}{2}}(\frac{D-k}{2})!}
M^{D-k}
\Bigl\{
\log\left(\frac{ML_k}{4\pi}\right)-\frac{1}{2}\left(\psi(1)+\psi(\tfrac{D-k}{2}+1)\right)
\Bigr\}\notag\\
&\hspace{5mm}+
\frac{(-1)(-1)^{\frac{D-k}{2}}}{2^{D-k}\pi^{\frac{D-k}{2}}}M^{D-k}\sum_{\ell=1}^{\infty}
\frac{(-1)^{\ell}\Gamma(2\ell +1)\zeta(2\ell +1)}{\Gamma(\ell+\tfrac{D-k}{2}+1)\Gamma(\ell+1)}\left(\frac{ML_k}{4\pi}\right)^{2\ell}
\notag\\
&\hspace{5mm}+
\frac{-1}{\pi^{\frac{D-k}{2}}L_k^{D-k}}\sum_{m=0}^{\frac{D-k}{2}-1}\frac{(-1)^m}{m!}\Gamma(\tfrac{D-k}{2}-m)\zeta(D-k -2m)
\left(\frac{ML_k}{2}\right)^{2m}.
\label{3.13}
\end{align}
The first term, the power-type nonanalytic term, arises from the residue at the 
single pole $t=\tfrac{1}{2}$ of $\zeta(2t)$ and the second term containing the logarithmic-type nonanalytic 
term originates from the residue at the double pole $t=0$ of the part $\Gamma(t-\tfrac{D-k}{2})\Gamma(t)$ 
in the integrand. The other two terms are analytic. We denote the nonanalytic terms in Eq.(\ref{3.13}) by
\begin{align}
F_{L_k}^{(1)D-k}(M)\Big|_{\text{n.a.}}^{\text{pow}}&
\equiv\frac{(-1)(-1)^{\frac{D-k}{2}}}{2^{\frac{D-k}{2}}\pi^{\frac{D-k-2}{2}}(D-k-1)!!}\frac{M^{D-k-1}}{L_k},
\label{3.14}
\\
F_{L_k}^{(1)D-k}(M)\Big|_{\text{n.a.}}^{\text{log}}&\equiv
\frac{(-1)(-1)^{\frac{D-k}{2}}}{2^{D-k}\pi^{\frac{D-k}{2}}(\frac{D-k}{2})!}
M^{D-k}\log\left(\frac{ML_k}{4\pi}\right).
\label{3.15}
\end{align}

For $(D, k)=({\text{even}}, {\text{odd}}), ({\text{odd}}, {\text{even}})$, we have
\begin{align}
F_{L_k}^{(1)D-k}(M)&=
\frac{(-1)^{\frac{D-k+1}{2}}}{2^{\frac{D-k+1}{2}}\pi^{\frac{D-k-1}{2}}(D-k)!!}M^{D-k}\notag\\
&\hspace{5mm}+
\frac{(-1)(-1)^{\frac{D-k-1}{2}}}{2^{D-k-1}\pi^{\frac{D-k-1}{2}}(\frac{D-k-1}{2})!}
\frac{M^{D-k-1}}{L_k}
\Bigl\{
-\log\left(ML_k\right)+\frac{1}{2}\left(\gamma_E+\psi(\tfrac{D-k-1}{2}+1)\right)
\Bigr\}\notag\\
&\hspace{5mm}+
\frac{-1}{\pi^{\frac{D-k}{2}}L_k^{D-k}}\sum_{\substack{m=0\\(m\neq \frac{D-k-1}{2})}}^{\infty}\frac{(-1)^m}{m!}
\Gamma(\tfrac{D-k}{2}-m)\zeta(D-k -2m)\left(\frac{ML_k}{2}\right)^{2m}.
\label{3.16}
\end{align}
The first term is the power-type nonanalytic term that comes from the residue at the 
single pole $t=0$ of $\Gamma(t)$ and the second term including the logarithmic-type nonanalytic 
term arises from the residue at the double pole $t=\frac{1}{2}$ of the part 
$\Gamma(t-\tfrac{D-k}{2})\zeta(2t)$ in the integrand. The 
last term is analytic. We represent the nonanalytic terms in Eq.(\ref{3.16}) as
\begin{align}
F_{L_k}^{(1)D-k}(M)\Big|_{\text{n.a.}}^{\text{pow}}&
\equiv \frac{(-1)^{\frac{D-k+1}{2}}}{2^{\frac{D-k+1}{2}}\pi^{\frac{D-k-1}{2}}(D-k)!!}M^{D-k},
\label{3.17}
\\
F_{L_k}^{(1)D-k}(M)\Big|_{\text{n.a.}}^{\text{log}}&\equiv
\frac{(-1)^{\frac{D-k-1}{2}}}{2^{D-k-1}\pi^{\frac{D-k-1}{2}}(\frac{D-k-1}{2})!}
\frac{M^{D-k-1}}{L_k}\log\left(ML_k\right).
\label{3.18}
\end{align}

We have evaluated the $t$ integral in Eq.(\ref{3.12}) by the residue theorem and have found the possible types of the 
nonanalytic terms in $F_{L_k}^{(1)D-k}(M)$ to be the power type or the logarithmic one. $F_{L_k}^{(1)D-k}(M)$ contains not 
only the nonanalytic terms but also the analytic ones. This implies that although $F_{L_k}^{(1)D-k}(M)$ is the contribution 
from the zero-mode sector through the process of applying the mode 
recombination formula, it incorporates more contributions in addition to the nonanalytic terms. We
will come back to this point again at the end of this section.

We are ready to calculate the nonanalytic terms in $V_{\text{eff}}$ according to Eq.(\ref{3.10}). 
We introduce the useful notation to study the nonanalytic term and discuss separately the cases of 
even and odd values of $D$ and $p+1$. For the case of $D=$ even, let us recall the notations defined in the 
paper II, which is convenient to discuss the power-type nonanalytic terms
\begin{align}
A_{2n+1}^{\text{pow}}&\equiv \frac{1}{\prod_{i=0}^{2n}L_i}F_{L_{2n+1}}^{(1)D-2n-1}(M)\Big|_{\rm n.a.}^{\text{pow}}=
\frac{(-1)^{\frac{D-2n}{2}}}{2^{\frac{D-2n}{2}}\pi^{\frac{D-2n-2}{2}}(D-2n-1)!!}\frac{M^{D-2n-1}}{L_0\cdots L_{2n}},
\label{3.19}
\\
B_{2n}^{\text{pow}}&\equiv \frac{1}{\prod_{i=0}^{2n-1}L_i}F_{L_{2n}}^{(1)D-2n}(M)\Big|_{\rm n.a.}^{\text{pow}}=
\frac{(-1)(-1)^{\frac{D-2n}{2}}}{2^{\frac{D-2n}{2}}\pi^{\frac{D-2n-2}{2}}(D-2n-1)!!}\frac{M^{D-2n-1}}{L_0\cdots L_{2n}},
\label{3.20}
\end{align}
where we have newly added the superscript ``pow'' in Eqs.(\ref{3.19}) and (\ref{3.20}). One immediately finds
\begin{align}
A_{2n+1}^{\text{pow}}+B_{2n}^{\text{pow}}=0\quad {\text {for}}\quad n=0,1,\cdots.
\label{3.21}
\end{align} 

We further introduce the notations for the logarithmic-type nonanalytic terms defined by
\begin{align}
A_{2n-1}^{\text{log}}&\equiv \frac{1}{\prod_{i=0}^{2n-2}L_i}F_{L_{2n-1}}^{(1)D-(2n-1)}(M)\Big|_{\rm n.a.}^{\text{log}}=
\frac{(-1)^{\frac{D-2n}{2}}}{2^{D-2n}\pi^{\frac{D-2n}{2}}(\frac{D-2n}{2})!}\frac{M^{D-2n}}{L_0\cdots L_{2n-1}}\log(ML_{2n-1}),
\label{3.22}\\
B_{2n}^{\text{log}}&\equiv \frac{1}{\prod_{i=0}^{2n-1}L_i}F_{L_{2n}}^{(1)D-2n}(M)\Big|_{\rm n.a.}^{\text{log}}=
\frac{(-1)(-1)^{\frac{D-2n}{2}}}{2^{D-2n}\pi^{\frac{D-2n}{2}}(\frac{D-2n}{2})!}\frac{M^{D-2n}}{L_0\cdots L_{2n-1}}
\log\left(\frac{ML_{2n}}{4\pi}\right),
\label{3.23}
\end{align}
where we have used Eqs.(\ref{3.18}) and (\ref{3.15}), respectively. We find 
\begin{align}
A_{2n-1}^{\text{log}}+B_{2n}^{\text{log}}=
\frac{(-1)^{\frac{D-2n}{2}}}{2^{D-2n}\pi^{\frac{D-2n}{2}}(\frac{D-2n}{2})!}\frac{M^{D-2n}}{L_0
\cdots L_{2n-1}}\log\left(\frac{ 4\pi L_{2n-1}}{L_{2n}}\right)
\quad {\text{for}}\quad n=0, 1, \cdots.
\label{3.24}
\end{align}

The $n=0$ contribution in Eq.(\ref{3.24}) should be read as
\begin{align}
A_{2n-1}^{\text{log}}\Big|_{n=0}+B_{2n}^{\text{log}}\Big|_{n=0}
=\frac{(-1)^{\frac{D}{2}}}{2^D\pi^{\frac{D}{2}}(\frac{D}{2})!}M^D\log\left(\frac{4\pi}{\mu L_0}\right),
\label{3.25}
\end{align}
where we have defined 
\begin{align}
A_{2n-1}^{\text{log}}\Big|_{n=0}&\equiv F^{(0)D}(M)\Big|_{\text n.a.}^{\text{log}}\equiv 
\frac{(-1)^{\frac{D}{2}}}{2^D\pi^{\frac{D}{2}}(\frac{D}{2})!}M^D\log\frac{M}{\mu},
\label{3.26}\\
B_{2n}^{\text{log}}\Big|_{n=0}&\equiv F_{L_0}^{(1)D}(M)\Big|_{\text n.a.}^{\text{log}}\equiv  
\frac{(-1)(-1)^{\frac{D}{2}}}{2^D\pi^{\frac{D}{2}}(\frac{D}{2})!}M^D\log\left(\frac{ML_0}{4\pi}\right).
\label{3.27}
\end{align}
Here, we have used Eq.(\ref{2.16}) with $f=0, {\cal N}=1$ for Eq.(\ref{3.26}), and Eq.(\ref{3.15}) with $k=0$ for 
Eq.(\ref{3.27}). The $\log M$-dependence in Eqs.(\ref{3.24}) and (\ref{3.25}) cancels out, leaving alternatively
the nonanalytic dependence on the scale ratios, $L_{2n-1}/L_{2n}$ and $1/\mu L_0$.
In particular, the cancelation of $\log M$ term in Eq.(\ref{3.25}) occurs between the zero-temperature 
contribution (\ref{3.26}) and the finite-temperature one (\ref{3.27}), which has been pointed out by Dolan and Jackiw in 
their paper\cite{dj}. This implies that the finite-temperature effective
potential in the case of no spatial $S_i^1$ direction $(p=0)$ does 
not have the $\log M$ dependence, but it still possesses the nonanalytic structure given by
the logarithmic term, $\log(1/\mu L_0)$. As we will see below, this is also the case for fermions.

Equipped with Eqs.(\ref{3.19}), (\ref{3.20}), (\ref{3.22}) and (\ref{3.23}), we can calculate the nonanalytic terms in the effective 
potential $V_{\text{eff}}$ by Eq.(\ref{3.10}) for $D=$ even.
\subsection{$(D, p+1)=$ (even, odd)}
The nonanalytic terms in $V_{\text{eff}}$ are written by
\begin{align}
V_{\text{eff}}\big|_{\text{n.a.}}&=\sum_{n=0}^{\frac{p}{2}}\Bigl\{A_{2n-1}^{\text{log}}+B_{2n}^{\text{log}}\Bigr\}+
\sum_{n=0}^{\frac{p}{2}-1}\Bigl\{A_{2n+1}^{\text{pow}}+B_{2n}^{\text{pow}}\Bigr\}+B_{2n}^{\text{pow}}\big|_{n=\frac{p}{2}}\notag\\
&=\sum_{n=0}^{\frac{p}{2}}
\frac{(-1)^{\frac{D-2n}{2}}}{2^{D-2n}\pi^{\frac{D-2n}{2}}(\frac{D-2n}{2})!}
\frac{M^{D-2n}}{L_0\cdots L_{2n-1}}\log\left(\frac{4\pi L_{2n-1}}{L_{2n}}\right)\notag\\
&\hspace{5mm}+\frac{(-1)(-1)^{\frac{D-p}{2}}}{2^{\frac{D-p}{2}}\pi^{\frac{D-p-2}{2}}(D-p-1)!!}\frac{M^{D-p-1}}{L_0\cdots L_p},
\label{3.28}
\end{align}
where we have used Eqs.(\ref{3.21}), (\ref{3.24}) and (\ref{3.20}) with $n=\frac{p}{2}$. The $n=0$ contribution 
in the sum over $n$ of Eq.(\ref{3.28}) is given by Eq.(\ref{3.25}). The effective potential has the power-type
nonanalytic term\cite{sakatake2022, sakatake2024} arising from $B_{2n}^{\text{pow}}|_{n=\frac{p}{2}}$ in this case.
The $\log M$-term cancels out and what remains is the nonanalytic term that depends on
the scale ratios, $L_{2n-1}/L_{2n}$ and $1/\mu L_0$.
\subsection{$(D, p+1)=$ (even, even)}
In this case the nonanalytic terms are given by
\begin{align}
V_{\text{eff}}\big|_{\text{n.a.}}&=\sum_{n=0}^{\frac{p-1}{2}}\Bigl\{A_{2n-1}^{\text{log}}+B_{2n}^{\text{log}}\Bigr\}
+A_{2n-1}^{\text{log}}\big|_{n=\frac{p+1}{2}}
+\sum_{n=0}^{\frac{p-1}{2}}\Bigl\{A_{2n+1}^{\text{pow}}+B_{2n}^{\text{pow}}\Bigr\}\notag\\
&=\sum_{n=0}^{\frac{p-1}{2}}\frac{(-1)^{\frac{D-2n}{2}}}{2^{D-2n}\pi^{\frac{D-2n}{2}}(\frac{D-2n}{2})!}
\frac{M^{D-2n}}{L_0\cdots L_{2n-1}}\log\left(\frac{4\pi L_{2n-1}}{L_{2n}}\right)\notag\\
&\hspace{5mm}+
\frac{(-1)^{\frac{D-p-1}{2}}}{2^{D-p-1}\pi^{\frac{D-p-1}{2}}(\frac{D-p-1}{2})!}
\frac{M^{D-p-1}}{L_0\cdots L_p}\log(ML_p),
\label{3.29}
\end{align}
 where we have used Eqs.(\ref{3.21}), (\ref{3.24}) and (\ref{3.22}) with $n=\frac{p+1}{2}$. Eq.(\ref{3.25}) should
 be understood as the $n=0$ contribution in the $n$ sum of Eq.(\ref{3.29}). The effective 
 potential possesses no power-type nonanalytic term, which is consistent with the results in the papers I and II.
On the other hand, the $\log M$ term arising from $A_{2n-1}^{\text{log}}|_{n=\frac{p+1}{2}}$ survives and 
other nonanalytic contributions, $\log(L_{2n-1}/L_{2n})$ and $\log(1/\mu L_0)$ also appear.

Let us proceed to study the case of $D=$ odd. We use the notations
defined in the paper II, which is useful to discuss the power-type nonanalytic terms
\begin{align}
C_{2n-1}^{\text{pow}}&\equiv \frac{1}{\prod_{i=0}^{2n-2}L_i}F_{L_{2n-1}}^{(1)D-(2n-1)}(M)\Big|_{\rm n.a.}^{\text{pow}}=
\frac{(-1)(-1)^{\frac{D-(2n-1)}{2}}}{2^{\frac{D-2n+1}{2}}\pi^{\frac{D-2n-1}{2}}(D-2n)!!}\frac{M^{D-2n}}{L_0\cdots L_{2n-1}},
\label{3.30}\\
D_{2n}^{\text{pow}}&\equiv \frac{1}{\prod_{i=0}^{2n-1}L_i}F_{L_{2n}}^{(1)D-2n}(M)\Big|_{\rm n.a.}^{\text{pow}}=
\frac{(-1)^{\frac{D-(2n-1)}{2}}}{2^{\frac{D-2n+1}{2}}\pi^{\frac{D-2n-1}{2}}(D-2n)!!}\frac{M^{D-2n}}{L_0\cdots L_{2n-1}},
\label{3.31}
\end{align}
where we have newly added the superscript ``pow'' in Eqs.(\ref{3.30}) and (\ref{3.31}). We 
see that the relation 
\begin{align}
C_{2n-1}^{\text{pow}}+D_{2n}^{\text{pow}}=0\quad {\text{for}}\quad n=0,1,\cdots
\label{3.32}
\end{align}
holds. In the case of $n=0$ in Eq.(\ref{3.32}), it should be understood as
\begin{align}
C_{2n-1}^{\text{pow}}\big|_{n=0}&\equiv F^{(0)D}(M)\Big|_{\text{n.a.}}^{\text{pow}}\equiv
\frac{(-1)(-1)^{\frac{D+1}{2}}}{2^{\frac{D+1}{2}}\pi^{\frac{D-1}{2}}D!!}M^D,
\label{3.33}\\
D_{2n}^{\text{pow}}\big|_{n=0}&\equiv F_{L_0}^{(1)D}(M)\Big|_{\text{n.a.}}^{\text{pow}}
=\frac{(-1)^{\frac{D+1}{2}}}{2^{\frac{D+1}{2}}\pi^{\frac{D-1}{2}}D!!}M^D,
\label{3.34}
\end{align}
where we have used Eq.(\ref{2.18}) with $f=0, {\cal N}=1$ and Eq.(\ref{3.17}) with $k=0$, respectively, which confirms 
the relation (\ref{3.32}) for $n=0$.

In addition to them, we further define the notations
\begin{align}
C_{2n+1}^{\text{log}}&\equiv \frac{1}{\prod_{i=0}^{2n}L_i}F_{L_{2n+1}}^{(1)D-(2n+1)}(M)\Big|_{\rm n.a.}^{\text{log}}=
\frac{(-1)(-1)^{\frac{D-2n-1}{2}}}{2^{D-2n-1}\pi^{\frac{D-2n-1}{2}}(\frac{D-2n-1}{2})!}\frac{M^{D-2n-1}}{L_0\cdots L_{2n}}
\log\left(\frac{ML_{2n+1}}{4\pi}\right),
\label{3.35}\\
D_{2n}^{\text{log}}&\equiv \frac{1}{\prod_{i=0}^{2n-1}L_i}F_{L_{2n}}^{(1)D-2n}(M)\Big|_{\rm n.a.}^{\text{log}}=
\frac{(-1)^{\frac{D-2n-1}{2}}}{2^{D-2n-1}\pi^{\frac{D-2n-1}{2}}(\frac{D-2n-1}{2})!}\frac{M^{D-2n-1}}{L_0\cdots L_{2n}}
\log(ML_{2n}),
\label{3.36}
\end{align}
where we have used Eqs.(\ref{3.15}) and (\ref{3.18}), respectively. Then, we find
\begin{align}
C_{2n+1}^{\text{log}}+D_{2n}^{\text{log}}
=\frac{(-1)^{\frac{D-2n-1}{2}}}{2^{D-2n-1}\pi^{\frac{D-2n-1}{2}}(\frac{D-2n-1}{2})!}\frac{M^{D-2n-1}}{L_0\cdots L_{2n}}
\log\left(\frac{ 4\pi L_{2n}}{L_{2n+1}}\right)
\quad {\text{for}}\quad n=0,1,\cdots~~.
\label{3.37}
\end{align}
The $\log M$ dependence drops out, leaving the nonanalytic dependence on the scale ratio $L_{2n}/L_{2n+1}$ instead.

With the help of Eqs.(\ref{3.30}), (\ref{3.31}), (\ref{3.35}) and (\ref{3.36}), one can calculate the nonanalytic terms in the
effective potential $V_{\text{eff}}$ according to Eq.(\ref{3.10}) for $D=$ odd.
\subsection{$(D, p+1)=$ (odd, odd)}
The nonanalytic terms in the effective potential $V_{\text{eff}}$ are given by
\begin{align}
V_{\text{eff}}\big|_{\text{n.a.}}&=\sum_{n=0}^{\frac{p}{2}-1}\Bigl\{C_{2n+1}^{\text{log}}+D_{2n}^{\text{log}}\Bigr\}
+D_{2n}^{\text{log}}\big|_{n=\frac{p}{2}}+
\sum_{n=0}^{\frac{p}{2}}\Bigl\{C_{2n-1}^{\text{pow}}+D_{2n}^{\text{pow}}\Bigr\}\notag\\
&=\sum_{n=0}^{\frac{p}{2}-1}
\frac{(-1)^{\frac{D-2n-1}{2}}}{2^{D-2n-1}\pi^{\frac{D-2n-1}{2}}(\frac{D-2n-1}{2})!}
\frac{M^{D-2n-1}}{L_0\cdots L_{2n}}\log\left(\frac{4\pi L_{2n}}{L_{2n+1}}\right)\notag\\
&\hspace{5mm}+
\frac{(-1)^{\frac{D-p-1}{2}}}{2^{D-p-1}\pi^{\frac{D-p-1}{2}}(\frac{D-p-1}{2})!}
\frac{M^{D-p-1}}{L_0\cdots L_p}\log(ML_p),
\label{3.38}
\end{align}
where we have used Eqs.(\ref{3.32}), (\ref{3.37}) and (\ref{3.36}) with $n=\frac{p}{2}$. We 
observe that there is no power-type nonanalytic term\cite{sakatake2022, sakatake2024}. The $\log M$ term
arising from $D_{2n}^{\text{log}}|_{n=\frac{p}{2}}$ survives and the $\log(L_{2n}/L_{2n+1})$ term also appears.  
\subsection{$(D, p+1)=$ (odd, even)}
In this case the nonanalytic terms in $V_{\text{eff}}$ are given by
\begin{align}
V_{\text{eff}}\big|_{\text{n.a.}}&=\sum_{n=0}^{\frac{p-1}{2}}
\Bigl\{C_{2n+1}^{\text{log}}+D_{2n}^{\text{log}}\Bigr\}+
\sum_{n=0}^{\frac{p-1}{2}}\Bigl\{C_{2n-1}^{\text{pow}}+D_{2n}^{\text{pow}}\Bigr\}
+C_{2n-1}^{\text{pow}}\big|_{n=\frac{p+1}{2}}
\notag\\
&=\sum_{n=0}^{\frac{p-1}{2}}
\frac{(-1)^{\frac{D-2n-1}{2}}}{2^{D-2n-1}\pi^{\frac{D-2n-1}{2}}(\frac{D-2n-1}{2})!}
\frac{M^{D-2n-1}}{L_0\cdots L_{2n}}\log\left(\frac{4\pi L_{2n}}{L_{2n+1}}\right)\notag\\
&\hspace{5mm}+
\frac{(-1)(-1)^{\frac{D-p}{2}}}{2^{\frac{D-p}{2}}\pi^{\frac{D-p-2}{2}}(D-p-1)!!}\frac{M^{D-p-1}}{L_0\cdots L_p},
\label{3.39}
\end{align}
where we have used Eqs.(\ref{3.32}), (\ref{3.37}) and (\ref{3.30}) with $n=\frac{p+1}{2}$. 
$V_{\text{eff}}$ has the power-type nonanalytic term arising from $C_{2n-1}^{\text{pow}}|_{n=\frac{p+1}{2}}$
\cite{sakatake2022, sakatake2024}. The $\log M$ dependence cancels out and 
the nonanalytic dependence on the scale ratio $L_{2n}/L_{2n+1}$ emerges instead.
Let us note that the overall sign $(-1)^{(D-p-1)/2}$ in the $\log M$ term
in Eqs.(\ref{3.29}) and (\ref{3.38}) takes $(D-p-1)/2=1~(-1)$ for $D-p-1=4m~(4m+2)$ with $m=1,2,\cdots$.

We obtained all the nonanalytic terms that cannot be expressed in the positive powers 
of $M^2$ in the effective potential $V_{\text{eff}}$ for the case of the scalar field satisfying the
periodic boundary condition for the spatial $S_i^1~(i=1,\cdots, p)$ direction. The effective potential 
has two types of nonanalytic terms: power type and logarithmic one. We demonstrated that 
the appearance of these terms is highly characteristic, that is, the power-type nonanalytic 
term exists in the case of $D-(p+1)=$ odd, while the logarithmic-type one
does in the case of $D-(p+1)=$ even. The two types never appear simultaneously. 
We summarize the results in Table $1$.
\begin{table}[ht]
\begin{center}
\begin{tabular}{l|c|c|c|c}
$(D, p+1)$                  & 	(even, odd) & (even, even)   &  (odd, odd)      & (odd, even)   \\ \hline\hline
power-type 	&       (3.28)     &   {\footnotesize non}           &{\footnotesize non}   & (3.39)\\
$\log M$-type 	&      {\footnotesize non}                 &(3.29)  &    (3.38)                     &  {\footnotesize non} \\\hline
\end{tabular}
\end{center}
\caption{Nonanalytic structure of the effective potential on the 
spacetime, $S_{\tau}^1\times R^{D-(p+1)}\times \prod_{i=1}^pS_i^1$ for the scalar field with the 
periodic boundary condition for the $S_i^1~(i=1,\cdots, p)$ direction. The ``non'' in the table means
that there is no corresponding type of nonanalytic terms. 
The numbers in the parentheses stand for the equation numbers in the text.}
\end{table}

It may be appropriate to comment on the nonanalytic terms, which have not mentioned in the
paper II. The zero modes $(n_0=n_1=\cdots= n_p=0)$ are crucial for the existence of the nonanalytic terms
and in fact the second term $F_{L_k}^{(1)D-k}(M)$ in Eq.(\ref{3.10}) containing the nonanalytic terms 
arises from the zero-mode sector through the process of applying the mode recombination formula. Since 
the infrared behavior of the theory may be dominated by the zero mode, such an infrared nonanalyticity
is expected to be obtained directly by the contribution from the zero mode. We show that this is
indeed the case below. By 
setting $n_0=n_1=\cdots=n_p=0, \eta_0=\eta_1=\cdots =\eta_p=0$ and $f=0, {\cal N}=1$ in Eq.(\ref{2.4}), we have
\begin{align}
V_{\text{eff}}^{\text{zero mode}}
=\frac{1}{2}\frac{1}{L_0\cdots L_p}\int \frac{d^{D-(p+1)}p_E}{(2\pi)^{D-(p+1)}} \log\left(p_E^2+M^2\right).
\label{3.40}
\end{align}
We follow the prescription to calculate the effective potential in section $2$ and immediately obtain 
\begin{align}
V_{\text{eff}}^{\text{zero mode}}=
-\frac{\pi^{\frac{D-(p+1)}{2}}}{2(2\pi)^{D-(p+1)}}\Gamma\left(-\tfrac{D-(p+1)}{2}\right)
\frac{M^{D-p-1}}{L_0\cdots L_p}.
\label{3.41}
\end{align}

For $D-(p+1)=$ odd, by using Eq.(\ref{2.17}), we find 
\begin{align}
V_{\text{eff}}^{\text{zero mode}}\Big|_{D-(p+1)={\text{odd}}}=
\frac{(-1)(-1)^{\frac{D-p}{2}}}{2^{\frac{D-p}{2}}\pi^{\frac{D-p-2}{2}}(D-p-1)!!}
\frac{M^{D-p-1}}{L_0\cdots L_p}.
\label{3.42}
\end{align}
This is the same with the second terms in Eqs.(\ref{3.28}) and (\ref{3.39}).
On the other hand, for $D-(p+1)=$ even, the gamma
function in Eq.(\ref{3.41}) has the pole and we employ, for example, the dimensional regularization and 
the $\overline{MS}$ renormalization to yield
\begin{align}
V_{\text{eff}}^{\text{zero mode}}\Big|_{D-(p+1)={\text{even}}}=
\frac{(-1)^{\frac{D-p-1}{2}}}{2^{D-p-1}\pi^{\frac{D-p-1}{2}}(\frac{D-p-1}{2})!}
\frac{M^{D-p-1}}{L_0\cdots L_p}\log\frac{M}{\mu},
\label{3.43}
\end{align}
where the scale $\mu$ is a set of possible renormalization conditions. This is equivalent to 
the second terms in Eqs.(\ref{3.29}) and (\ref{3.38}) if we identify the scale $\mu$ as $L_p^{-1}$.

The above calculation shows that the nonanalytic terms are simply obtained by the contribution from only the zero mode
in Eq.(\ref{2.4}). 
Even though $F_{L_k}^{(1)D-k}(M)$ that potentially contains the nonanalytic terms also comes from the zero-mode sector through
the process of applying the mode recombination formula, due the existence of the winding mode
$m_k$ in  $F_{L_k}^{(1)D-k}(M)$, it includes additional quantum effects to result
the third and fourth terms in Eq.(\ref{3.13}) (or the third term in Eq.(\ref{3.16})) in addition to the nonanalytic 
 terms. Therefore, our approach for studying the nonanalytic terms leads to a larger contribution 
 than the approach that takes into account only the zero mode.
%
%
%
\section{Nonanalytic terms for fermion field with arbitrary boundary condition}
%
%
%
\subsection{New form of $V_{\text{eff}}^f$}
Let us study the nonanalytic structure of the effective potential $V_{\text{eff}}^f$ in the case of
fermion $(f=1, \eta_0=\frac{1}{2})$ with arbitrary boundary condition for the spatial $S^1_i~(i=1,\cdots, p)$ direction.
It is convenient to use the expression(\ref{2.24}) and we consider the 
second term whose integral representation in the complex plane is given by Eq.(\ref{2.26}) 
\begin{align}
&\frac{1}{L_0}\sum_{n_0=-\infty}^{\infty}F_{L_{i_1},\cdots, L_{i_n}}^{(n)(D-1, f)}(M_{(0)})\notag\\
&\hspace{5mm}
={\cal N}\frac{2^n\pi^{\frac{D-1}{2}}}{L_0^D}
\frac{1}{4\pi i}
\int_{c-i\infty}^{c+i\infty}dt~\Gamma(t-\tfrac{D-1}{2})\sum_{n_0=-\infty}^{\infty}
\Biggl\{
\left(n_0+\frac{1}{2}\right)^2 +\left(\frac{ML_0}{2\pi}\right)^2
\Biggr\}^{-t+\frac{D-1}{2}}\notag\\
&\hspace{15mm}\times
\left(\frac{\pi}{L_0}\right)^{-2t}S^{(n)}(t; L_{i_1},\cdots, L_{i_n}).
\label{4.1}
\end{align}
Hereafter, we assign the superscript ``{\it f} '' in the relevant quantity like $F_{L_{i_1},\cdots, L_{i_n}}^{(n)(D-1, f)}(M_{(0)})$
in order to make it clear that we are considering the case of fermions.

The mode sum over $n_0$ has the analytical extension\cite{sakatake2024, elizalde}
\begin{align}
&\Gamma(t)\sum_{n_0=-\infty}^{\infty}
\Biggl\{
\left(n_0+\half\right)^2+z^2\Biggr\}^{-t}\notag\\
&
=\sqrt{\pi}\frac{\Gamma(t-\half)}{z^{2(t-\half)}}
-\frac{4\pi^{2t}}{\sqrt{\pi}}\frac{1}{4\pi i}\int_{c_1-i\infty}^{c_1+i\infty}dt_1~\Gamma(t_1-t+\tfrac{1}{2})\zeta(2t_1-2t+1)\Gamma(t_1)(\pi z)^{-2t_1}\notag\\
&\hspace{2.8cm}
+\frac{4(2\pi)^{2t}}{\sqrt{\pi}}\frac{1}{4\pi i}\int_{c_1-i\infty}^{c_1+i\infty}dt_1~\Gamma(t_1-t+\tfrac{1}{2})\zeta(2t_1-2t+1)\Gamma(t_1)(2\pi z)^{-2t_1}
\notag\\
&
=\sqrt{\pi}\frac{\Gamma(t-\half)}{z^{2(t-\half)}}
-\frac{4\pi^{2t}}{\sqrt{\pi}}\frac{1}{4\pi i}\int_{c_1-i\infty}^{c_1+i\infty}dt_1~\Gamma(t_1-t+\tfrac{1}{2})\eta(2t_1-2t+1)\Gamma(t_1)(\pi z)^{-2t_1},
\label{4.2}
\end{align}
where we have defined the so-called eta function
\begin{align}
\eta(t)\equiv (1-2^{1-t})\zeta(t).
\label{4.3}
\end{align}
The characteristic point in Eq.(\ref{4.2}), compared with Eq.(\ref{3.3}), is that the eta function appears instead of the zeta 
function and there is no counterpart corresponding to the first term arising from the zero mode $(n_0=0)$. These 
come from the fact that in the case of fermions the zero mode is removed due to the antiperiodic 
boundary condition for the Euclidean time direction followed from quantum statistics. After 
changing the variable $\bar t=t-\frac{D-1}{2}$ in Eq.(\ref{4.1}) and inserting Eq.(\ref{4.2}) into it, we obtain
\begin{align}
&\frac{1}{L_0}\sum_{n_0=-\infty}^{\infty}F_{L_{i_1},\cdots, L_{i_n}}^{(n)(D-1, f)}(M_{(0)})\notag\\
&=\frac{1}{L_0}{\cal N}\frac{2^n}{\pi^{\frac{D-1}{2}}}\frac{1}{4\pi i}\int_{c-\frac{D-1}{2}-i\infty}^{c-\frac{D-1}{2}+i\infty}dt~\Biggl\{
\sqrt{\pi}\Gamma(t-\tfrac{1}{2})\left(\frac{ML_0}{2\pi}\right)^{-2(t-\half)}\notag\\
&\hspace{5mm}
-\frac{4\pi^{2t}}{\sqrt{\pi}}\frac{1}{4\pi i}\int_{c_1-i\infty}^{c_1+i\infty}dt_1~\Gamma(t_1-t+\tfrac{1}{2})\eta(2t_1-2t+1)\Gamma(t_1)
\left(\frac{ML_0}{2}\right)^{-2t_1}
\Biggr\}\notag\\
&\hspace{10mm}\times
\left(\frac{\pi}{L_0}\right)^{-2t}S^{(n)}(t+\tfrac{D-1}{2}; L_{i_1},\cdots, L_{i_n}),
\label{4.4}
\end{align}
where we have denoted $\bar t$ as $t$ again.

We deform the $t_1$ integration path in Eq.(\ref{4.4}) so that it encloses all the poles of the integrand and then evaluate the 
$t_1$ integral by the residue theorem. For this to be possible, as discussed in Appendix A, the condition
\begin{align}
0 < \frac{ML_0}{\pi} < 1
\label{4.5}
\end{align}
must be satisfied. Among the poles $t_1=t-\half -\ell~(\ell=0,1,\cdots)$ of $\Gamma(t_1-t+\tfrac{1}{2})$, only the pole
$t_1=t-\tfrac{1}{2}$ is relevant because of $\eta(-2\ell)=0~(\ell=1,2,\cdots)$, which always follows from the 
combination $\Gamma(t_1-t+\tfrac{1}{2})\eta(2t_1-2t+1)$ and is frequently used throughout 
our discussions. The residue at the pole $t_1=t-\tfrac{1}{2}$ with $\eta(0)=1/2$ cancels the first term in Eq.(4.4). 
The eta function does not possesses any pole unlike the zeta function.
We end up with the contribution from the pole $t_1=-\bar n~(\bar n=0,1,\cdots)$ of $\Gamma(t_1)$, so that 
Eq.(\ref{4.4}) becomes 
\begin{align}
&\frac{1}{L_0}\sum_{n_0=-\infty}^{\infty}F_{L_{i_1},\cdots, L_{i_n}}^{(n)(D-1, f)}(M_{(0)})\notag\\
&=-{\cal N}\frac{2^n}{\pi^{\frac{D}{2}}L_0}\frac{1}{2\pi i}\int_{c-\frac{D-1}{2} -i\infty}^{c-\frac{D-1}{2}+i\infty}dt~
\sum_{\bar n=0}^{\infty}\frac{(-1)^{\bar n}}{\bar n!}\left(\frac{ML_0}{2}\right)^{2\bar n}
\Gamma(-t-\bar n +\tfrac{1}{2})
\eta(-2t-2\bar n  +1)
\notag\\
&\hspace{5mm}\times \left(\frac{1}{L_0}\right)^{-2t}S^{(n)}(t+\tfrac{D-1}{2}; L_{i_1},\cdots, L_{i_n}).
\label{4.6}
\end{align}
There may be the poles in the multiple mode summations $S^{(n)}(t+\tfrac{D-1}{2};L_{i_1},\cdots, L_{i_n})$
that contribute to the above $t$ integral. It does not, however, affect 
the $(M^2)^{\bar n}$ dependence due to the nonexistence of $M$ in the summations. 
We observe that all the terms in Eq.(\ref{4.6}) are analytic and can be expressed in the positive powers of $M^2$.
Hence, there exist no nonanalytic terms in Eq.(\ref{4.6}).

Let us next study the first term in Eq.(\ref{2.24}), whose explicit form is given by Eq.(\ref{2.28})
 with $f=1$ and $\eta_0=\tfrac{1}{2}$
\begin{align}
\frac{1}{L_0}\sum_{n_0=-\infty}^{\infty}F^{(0)(D-1, f)}(M_{(0)})
={\cal N}\frac{\pi^{\frac{D-1}{2}}}{2L_0^D}\Gamma(-\tfrac{D-1}{2})\sum_{n_0=-\infty}^{\infty}
\Biggl\{
\left(n_0+\half\right)^2+\left(\frac{ML_0}{2\pi}\right)^2\Biggr\}^{\frac{D-1}{2}}.
\label{4.7}
\end{align}
The analytical extension for the mode sum over $n_0$ is given by 
setting $t=-\frac{D-1}{2}, z=\tfrac{ML_0}{2\pi}$ in Eq.(\ref{4.2}) as
\begin{align}
\frac{1}{L_0}\sum_{n_0=-\infty}^{\infty}F^{(0)(D-1, f)}(M_{(0)})&=
{\cal N}\frac{\pi^{\frac{D}{2}}}{2(2\pi )^D}\Gamma(-\tfrac{D}{2})M^D\notag\\
&\hspace{3mm}-{\cal N}\frac{1}{L_0^D\pi^{\frac{D}{2}}}\frac{1}{2\pi i}\int_{c_1-i\infty}^{c_1+i\infty}
dt_1~\Gamma(t_1+\tfrac{D}{2})\eta(2t_1+D)\Gamma(t_1)\left(\frac{ML_0}{2}\right)^{-2t_1}.
\label{4.8}
\end{align}

The first term in Eq.(\ref{4.8}) is precisely the $F^{(0)(D, f)}(M)$ from Eq.(\ref{2.15}) with $f=1$. 
On the other hand, the second term in Eq.(\ref{4.8}) is equivalent to $F_{L_0}^{(1)(D, f)}(M)$. To see 
this, we set $f=1, n=1, \eta_0=\tfrac{1}{2}$ in Eq.(\ref{2.14}), identifying $L_{i_1}$ as $L_0$ and use Eq.(\ref{2.25}) together with the 
definition of the eta function $\eta(z)=\sum_{m=1}^{\infty}(-1)^{m-1}m^{-z}$. Then, we have 
\begin{align}
F_{L_0}^{(1)(D, f)}(M)=-{\cal N}\frac{1}{(2\pi)^{\frac{D}{2}}}\left(\frac{M^2}{2}\right)^{\frac{D}{2}}\frac{1}{2\pi i}\int_{c-i\infty}^{c+i\infty}
dt~\Gamma(t-\tfrac{D}{2})\Gamma(t)\eta(2t)\left(\frac{ML_0}{2}\right)^{-2t}.
\label{4.9}
\end{align}
And if we change the variable $t_1=t-\tfrac{D}{2}$ in Eq.(\ref{4.9}), we arrive at the second term in Eq.(\ref{4.8}).

Then, we finally obtain from Eq.(\ref{2.24}) with Eqs.(\ref{4.6}) and (\ref{4.8}) that 
\begin{align}
V_{\text{eff}}^f&={\cal N}\frac{\pi^{\frac{D}{2}}}{2(2\pi )^D}\Gamma(-\tfrac{D}{2})M^D\notag\\
&\hspace{5mm}-{\cal N}\frac{1}{L_0^D\pi^{\frac{D}{2}}}\frac{1}{2\pi i}\int_{c_1-i\infty}^{c_1+i\infty}
dt_1~\Gamma(t_1+\tfrac{D}{2})\eta(2t_1+D)\Gamma(t_1)\left(\frac{ML_0}{2}\right)^{-2t_1}\notag\\
&\hspace{5mm}
-\sum_{n=1}^p\sum_{1\leq i_1<\cdots <i_n\leq p}
{\cal N}\frac{2^n}{\pi^{\frac{D}{2}}L_0}
\sum_{\bar n=0}^{\infty}\frac{(-1)^{\bar n}}{\bar n!}\left(\frac{ML_0}{2}\right)^{2\bar n}
\frac{1}{2\pi i}\int_{c-\frac{D-1}{2} -i\infty}^{c-\frac{D-1}{2}+i\infty}dt~
\notag\\
&\hspace{10mm}\times 
\Gamma( -t-\bar n+\tfrac{1}{2})
\eta(-2t-2\bar n  +1)\left(\frac{1}{L_0}\right)^{-2t}S^{(n)}(t+\tfrac{D-1}{2}; L_{i_1},\cdots, L_{i_n}).
\label{4.10}
\end{align}
We obtain an important expression for the effective potential in the case of the fermion with
arbitrary boundary condition for the spatial $S_i^1~(i=1,\cdots, p)$ direction. As stated 
below Eq.(\ref{4.6}), the third term in Eq.(\ref{4.10}) is strictly analytic with respect to $M$. 
We observe that the mode recombination formula clearly separates the effective potential $V_{\text{eff}}^f$ 
into a part that contains the nonanalytic terms and a part that is entirely analytic.
Hence, we conclude that the nonanalytic terms, if present, can arise from the first and second 
terms in Eq.(\ref{4.10}) and we write the nonanalytic terms in $V_{\text{eff}}^f$ by
\begin{align}
V_{\text{eff}}^f\Big|_{\text{n.a.}}=F^{(0)(D, f)}(M)\Big|_{\text{n.a.}}+ F_{L_0}^{(1)(D, f)}(M)\Big|_{\text{n.a.}}.
\label{4.11}
\end{align}
\subsection{Nonanalytic terms}
We have already calculated the first term in Eq.(\ref{4.10}) in section $2$ and have found the power-type or the logarithmic-type
nonanalytic terms. For $D=$ even, it is given by Eq.(\ref{2.16}) with $f=1$
\begin{align}
F^{(0)(D, f)}(M)\Big|_{n.a.}^{\text{log}}
&=
-{\cal N}\frac{(-1)^{\frac{D}{2}}}{2^D\pi^{\frac{D}{2}}(\frac{D}{2})!}M^D
\log\frac{M}{\mu},
\label{4.12}
\end{align}
while for $D=$ odd, it is done by Eq.(\ref{2.18}) with $f=1$
\begin{align}
F^{(0)(D, f)}(M)\Big|_{n.a.}^{\text{pow}}
&=
{\cal N}\frac{(-1)^{\frac{D+1}{2}}}{2^{\frac{D+1}{2}}\pi^{\frac{D-1}{2}}D!!}M^D.
\label{4.13}
\end{align}

Let us obtain the nonanalytic terms in the second term of Eq.(\ref{4.10}), which is $F_{L_0}^{(1)(D, f)}(M)$. We deform the 
contour in the complex $t_1$ plane to encircle all the poles of the integrand and evaluate the integral by 
the residue theorem. One needs the condition (\ref{4.5}) again for this to be possible, which will be 
derived in Appendix A. It is appropriate to compute the integral
separately for even and odd $D$.  The details of the calculation are given in Appendix C.
We present the result here. For $D=$ even with Eqs.(\ref{c.2}), (\ref{c.3}) and (\ref{c.5}), we have
\begin{align}
F_{L_0}^{(1)(D,f)}(M)&={\cal N}\frac{(-1)^{\frac{D}{2}}}{2^D\pi^{\frac{D}{2}}\left(\frac{D}{2}\right)!}
M^D 
\Biggl\{\log\left(\frac{ML_0}{\pi}\right)-\frac{1}{2}\left(\psi(1)+\psi(\tfrac{D}{2}+1)\right)
\Biggr\}\notag\\
&\hspace{5mm}-{\cal N}\frac{(-1)^{\frac{D}{2}}}{2^D\pi^{\frac{D}{2}}}M^D
\sum_{\ell=1}^{\infty}\frac{(-1)^{\ell}\Gamma(2\ell +1)(1-2^{1+2\ell})}{\Gamma(\ell+\tfrac{D}{2}+1)\Gamma(\ell+1)}
\zeta(2\ell +1)\left(\frac{ML_0}{4\pi}\right)^{2\ell}\notag\\
&\hspace{5mm}-{\cal N}\frac{1}{\pi^{\frac{D}{2}}L_0^D}\sum_{m=0}^{\frac{D}{2}-1}
\frac{(-1)^m}{m!}\Gamma(\tfrac{D}{2}-m)\eta(D-2m)\left(\frac{ML_0}{2}\right)^{2m}.
\label{4.14}
\end{align}
The first term in Eq.(\ref{4.14}) containing the logarithmic-type nonanalytic term
arises from the residue at the double pole $t_1=-\tfrac{D}{2}$ of the part $\Gamma(t_1+\tfrac{D}{2})\Gamma(t_1)$ in the
integrand. There is no power-type nonanalytic term in Eq.(\ref{4.14}), which reflects the fact that
the eta function $\eta(2t_1+D)$ does not possess the pole unlike the zeta function.

From Eqs.(\ref{4.12}) and (\ref{4.14}), there is no power-type nonanalytic term 
in the effective potential, which is consistent with the result in the paper II. We have only the logarithmic-type nonanalytic term
in $F_{L_0}^{(1)(D, f)}(M)$
\begin{align}
F_{L_0}^{(1)(D,f)}(M)\Big|_{\text{n.a.}}^{\text{log}}={\cal N}\frac{(-1)^{\frac{D}{2}}}{2^D\pi^{\frac{D}{2}}\left(\frac{D}{2}\right)!}
M^D \log\left(\frac{ML_0}{\pi}\right),
\label{4.15}
\end{align}
so that Eq.(\ref{4.11}) becomes
\begin{align}
V_{\text{eff}}^f\Big|_{\text{n.a.}}=F^{(0)(D, f)}(M)\Big|_{n.a.}^{\text{log}}+F_{L_0}^{(1)(D,f)}(M)\Big|_{\text{n.a.}}^{\text{log}}
={\cal N}\frac{(-1)^{\frac{D}{2}}}{2^D\pi^{\frac{D}{2}}\left(\frac{D}{2}\right)!}
M^D \log\left(\frac{\mu L_0}{\pi}\right).
\label{4.16}
\end{align}
The $\log M$ dependence finally cancels out, leaving alternatively the nonanalytic dependence on the scale $1/\mu L_0$
\footnote{This is the same situation as Eq.(\ref{3.25}), in which the $\log M$ dependence drops out.}.

Let us proceed to study the second term in Eq.(\ref{4.10}) for $D=$ odd. As discussed 
in Appendix C, the integrand has no double pole in this case, so that there arises 
no logarithmic-type nonanalytic terms. We obtain (see Eq.(\ref{c.6}))
\begin{align}
F_{L_0}^{(1)(D,f)}(M)&=
-{\cal N}\frac{(-1)^{\frac{D+1}{2}}}{2^{\frac{D+1}{2}}\pi^{\frac{D-1}{2}}D!!}M^D
\notag\\
&\hspace{5mm}-{\cal N}\frac{1}{\pi^{\frac{D}{2}}L_0^D}\sum_{\ell=0}^{\infty}
\frac{(-1)^{\ell}}{\ell !}\Gamma(\tfrac{D}{2}-\ell)\eta(D-2\ell)\left(\frac{ML_0}{2}\right)^{2\ell}.
\label{4.17}
\end{align}
The first term is the contribution from the single pole $t_1=-\tfrac{D}{2}$ of $\Gamma(t_1+\tfrac{D}{2})$ and 
the second one is the one from the pole $t_1=-\ell~(\ell=0,1,\cdots)$ of $\Gamma(t_1)$.
As expected, there is no logarithmic-type nonanalytic term in Eq.(\ref{4.17}). We see 
that the first term, the power-type nonanalytic term, cancels out Eq.(\ref{4.13}) to 
reproduce the result in the paper II. Therefore, the effective 
potential does not possess the nonanalytic terms
\begin{align}
V_{\text{eff}}^f\Big|_{\text{n.a.}}=0 \quad\text{for}\quad D=\text{odd}.
\label{4.18}
\end{align}

We conclude that in the case of the fermion with arbitrary boundary condition for the
spatial $S^1_i~(i=1,\cdots, p)$ direction, the effective potential does not have any nonanalytic 
term with respect to $M$. We summarize the results in Table $2$. This 
is quite different from the case of the scalar with the periodic boundary condition
for the spatial $S^1_i$ direction.
\\
\begin{table}[ht]
\begin{center}
\begin{tabular}{c|c|c}
              & $D=$ even &  $D=$ odd  \\ \hline\hline
power-type 	&       {\footnotesize non}     &   {\footnotesize non}          \\
$\log M$-type 	&      {\footnotesize non}      &  {\footnotesize non}     \\\hline
\end{tabular}
\end{center}
\caption{Nonanalytic structure of the effective potential $V_{\text{eff}}^f$ on the 
spacetime, $S_{\tau}^1\times R^{D-(p+1)}\times \prod_{i=1}^pS_i^1$ for the fermion field satisfying the arbitrary boundary
condition for the spatial $S^1_i~(i=1,\cdots, p)$ direction. The ``non'' in the table means
that the effective potential contains no nonanalytic term of the corresponding type.}
\end{table}

If we consider the scalar field with some of the boundary condition for a spatial $S_i^1$ direction being antiperiodic, the 
effective potential in this case essentially has the same with that in the case of the fermion 
by regarding the spatial direction that has the antiperiodic boundary condition as the Euclidean time direction.
Hence, the results in the case of the fermion immediately imply the conclusion that 
there are no nonanalytic terms if the scalar field obeys at least one antiperiodic 
boundary condition for the spatial $S_i^1$ direction.

%
%
%
\section{Conclusions and discussions}
%
We have investigated the nonanalytic terms in the one-loop finite-temperature effective 
potentials of the scalar and fermion fields on a $D$-dimensional 
spacetime, $S_{\tau}^1\times R^{D-(p+1)}\times \prod_{i=1}^p S_i^1$. 
These are the terms that cannot be expressed as the positive powers of the 
field-dependent mass squared. In our analysis we have made the extensive 
use of the mode recombination formula developed in the paper II, which 
can be applicable to fermions, scalars and arbitrary boundary conditions for the 
spatial direction $S_i^1~(i=1,\cdots, p)$. We recast the effective potential into the remarkable 
expressions, Eqs.(\ref{3.9}) and (\ref{4.10}) that 
have not been obtained previously. The rewritten expression clearly separates the analytic terms from all other 
ones, thereby making it transparent where the nonanalytic terms of interest arise.

The relevant quantities such as $F_{L_k}^{(1)D-k}(M), F_{L_0}^{(1)(D, f)}(M)$ 
that can potentially give rise to the nonanalytic terms in  the effective potential 
are initially expressed in terms of the modified Bessel functions of the second kind accompanied by the 
mode summations. By employing the integral representations of the function together with the analytical 
extension of the mode summations, we rewrite them as the contour integrals in the complex plane. We evaluate the integral by the 
residue theorem, picking up all infinitely many poles contained in the integrand. 
This procedure enables us to obtain the complete set of terms that appear in the effective potential. As a result, we find
 that the possible nonanalytic terms with respect to $M$ in the effective potential are of power type or logarithmic one.

The logarithmic-type nonanalytic terms newly analyzed in the present paper are found to arise from the double pole
in the integrand, in contrast to the power-type nonanalytic terms that originate from the single pole as shown in the papers I and II.
In the case of the scalar field obeying the periodic boundary condition for the spatial $S_i^1~(i=1,\cdots, p)$ direction, we have
determined all the nonanalytic terms in the effective potential by classifying the cases according to  
whether the spacetime dimension, $D$ and the total number of $S^1_i$, $p+1$ are even or odd. We have clarified
the significant nonanalytic structure as summarized in Table $1$: whenever the power type is 
present(absent), the logarithmic type is absent(present). We have also studied  the case of 
the fermion with arbitrary boundary condition for the spatial $S^1_i$ direction. The absence of the zero mode 
due to the antiperiodic boundary condition following from the quantum statistics is crucial for 
the nonanalytic structure and we have found that neither
of the two types appears in the effective potential regardless of the values of $D$ and $p+1$. 
The results in the case  of the fermion have immediately led to the conclusion that there exists 
no nonanalytic term in the case of the scalar satisfying the antiperiodic boundary
condition for at least one spatial $S_i^1$ direction.\footnote{We also discussed the nonanalytic terms in 
the case of a higher dimensional gauge field on $S_{\tau}^1\times R^{D-(p+1)}\times \prod_{i=1}^pS_i^1$ in 
the paper II. The results concerning with the nonanalytic terms obtained in this paper also hold for the gauge field.
The component gauge field for the spatial $S_i^1$ direction can acquire the vacuum expectation value to
yield the fractional values of $\eta_i$ in the boundary condition (\ref{2.3}). It may be important to study the 
analytical extension in that case. 
}

We have also shown that the nonanalytic terms (\ref{3.42}) and (\ref{3.43}) can be obtained directly from the 
contribution of the zero mode alone. In our formulation based on the mode recombination 
formula, $F_{L_k}^{(1)D-k}(M)$ that potentially contains the nonanalytic terms also arises from the zero-mode sector 
in the formula. However, thanks to the winding mode $m_k$ in $F_{L_k}^{(1)D-k}(M)$, it 
yields the larger contributions, the third and fourth terms in Eq.(\ref{3.13}) (or the
third term in Eq.(\ref{3.16})) in addition to the nonanalytic terms. Therefore, our formulation
provides the richer results that cannot be obtained by considering only the zero mode. 
Moreover, the origin and the mechanism responsible for the 
emergence of the nonanalytic terms are very clear in the formulation and it serves as powerful framework for 
analyzing the terms.

We have studied the nonanalytic terms that cannot be expressed by the positive powers of the field-dependent mass 
squared. In order to fully understand the nonanalytic structure of the effective potential, one needs to evaluate 
the complex $t$ integral in the third term of Eq.(\ref{3.9}) (or Eq.(\ref{4.10})). Even though the third term is 
entirely analytic in $M$, we expect that the
term contains nonanalytic contributions with power and/or logarithmic dependence 
on the scale $L_i~(i=0,\cdots, p)$ and/or their ratio. In fact, these types of the contributions already 
appear, for example, in the first term of Eq.(\ref{3.28}). It is of considerable interest to clarify the 
nonanalytic structure associated with the scale or the ratio, namely in connection with the phase-transition 
phenomena of the theory. A detailed investigation for fixed values of $p$ is currently 
in progress and will be reported in the near future.

We have obtained the nonanalytic terms in the effective potential under the condition (\ref{3.5}) (or (\ref{4.5})).
One could have new types of nonanalytic terms, which are different from the ones obtained in this 
paper, outside of the region of Eqs.(\ref{3.5}), (\ref{3.11}) and (\ref{4.5}). In this case, unlike the method adopted in this paper, it is necessary to 
evaluate (\ref{2.14}) directly by some means. The method is not known at present, but the problem is  highly challenging
and interesting.
%
%
%
%
%
%
%
\begin{center}
{\bf Acknowledgements}
\end{center}
This work is supported in part by Grants-in-Aid for Scientific 
Research [No.~18K03649 and No.~23K03416(M.S.)] from the Ministry of 
Education, Culture, Sports, Science and Technology (MEXT) in Japan.
\appendix
%
%
\section{Derivation of Eqs.(\ref{3.5}) and (\ref{4.5})}
Let us derive the conditions (\ref{3.5}) and (\ref{4.5}) in the text. We 
deform the contour in the $t_1$ plane of Eq.(\ref{3.4}) (or Eq.(\ref{4.4})) to 
encircle all the poles of the integrand and evaluate the integral by the residue theorem.
For this to be possible, the contribution from the large left semicircle must vanish, from which 
one obtains the conditions.

The relevant part in Eq.(\ref{3.4}) for the discussion is given by
\begin{align}
 \int_{c_1-i\infty}^{c_1+i\infty}dt_1~\Gamma(t_1-t+\tfrac{1}{2})\zeta(2t_1-2t +1)
 \Gamma(t_1)\left(\frac{ML_0}{2}\right)^{-2t_1}.
 \label{a.1}
 \end{align}
In order to examine the contribution from the large left semicircle
for the contour integral (\ref{a.1}), we study the asymptotic behavior of $\log\abs{I}$ in the large radius 
limit, where
 \begin{align}
 I\equiv \Gamma(t_1-t+\tfrac{1}{2})\zeta(2t_1-2t +1)\Gamma(t_1)\left(\frac{ML_0}{2}\right)^{-2t_1}.
\label{a.2}
\end{align}
To this end, we need to understand the asymptotic behavior of each function in the integrand.

The Stirling's formula is given by\cite{bromwich}
 \begin{align}
 \Gamma(\omega)\simeq \sqrt{2\pi}~\e^{-\omega}~\omega^{\omega -\frac{1}{2}} (1+\cdots ) 
 \quad \mbox{for}\quad \abs{\omega}\rightarrow \infty\quad \mbox{with}\quad  \abs{{\text {arg}}~\omega} < \pi,
 \label{a.3}
 \end{align}
 from which it follows that
 \begin{align}
\log\abs{\Gamma(\omega)}&\simeq {\text{Re}}~\Bigl\{
(\omega-\tfrac{1}{2})\log \omega -\omega +\frac{1}{2}\log(2\pi)+\cdots
\Bigr\}\notag\\
&=\left(\omega_R-\frac{1}{2}\right)\log\abs{\omega}-(\text{arg}~\omega)\omega_I -\omega_R+\cdots .
\label{a.4}
 \end{align}
Here we have denoted the real (imaginary) part of $\omega$ by $\omega_R(\omega_I)$.

The large left semicircle is parametrized by
\begin{align}
t_1=c+R\e^{i\varphi}\quad{\text{with}}\quad \frac{\pi}{2}\leq \varphi\leq \frac{3}{2}\pi,\quad R\rightarrow \infty,
\label{a.5}
\end{align}
 where $R>0, c\in\mathbb{R}$. The asymptotic 
 behavior\footnote{One should keep the limit $R\rightarrow \infty$ in mind when studying the asymptotic behavior in Appendix A.}
 of $\Gamma(t_1)$ in the limit $R\rightarrow \infty$ is obtained by using the Stirling's formula (\ref{a.3}) with $\omega$ 
 being identified with $t_1$. Taking into account $\text{arg}~t_1\simeq\text{arg}~\varphi~(\text{mod}~2\pi)$ in the 
limit $R\rightarrow \infty$, the range of validity of the formula (\ref{a.3}) is now 
given by $\abs{\text{arg}~\varphi} < \pi$, so that the range $\frac{\pi}{2}\leq \varphi\leq \frac{3}{2}\pi$ (especially $\varphi=\pi$)
lies outside of it.

In order to obtain the asymptotic behavior of the gamma function that is valid in the entire
range $\frac{\pi}{2}\leq \varphi\leq \frac{3}{2}\pi$, let us 
consider $\Gamma(\alpha -\beta t_1) $, where $\alpha\in \mathbb{C}, \beta >0$.
By identifying $\omega$ in Eq.(\ref{a.3}) with $\alpha-\beta t_1$, Eq.(\ref{a.4}) becomes 
\begin{align}
\log\abs{\Gamma(\alpha-\beta t_1)}\simeq 
-\beta R\cos\varphi \log(\beta R)+\beta R(\theta \sin\varphi +\cos\varphi)
+(\alpha_R-\beta c-\tfrac{1}{2})\log(\beta R)
\label{a.6}
\end{align}
in the limit $R\rightarrow\infty$, where we have denoted the real part 
of $\alpha$ by $\alpha_R$ and $\theta = \text{arg}~(\alpha-\beta t_1)$. 
Since Eq.(\ref{a.6}) is valid for $\abs{\theta}<\pi$ from Eq.(\ref{a.3}) and 
$\theta\simeq {\text {arg}}(-\e^{i\varphi})$ in the limit $R\rightarrow \infty$, Eq.(\ref{a.6}) is valid for
$\frac{\pi}{2}\leq \varphi\leq \frac{3}{2}\pi$ with $\theta\simeq\varphi-\pi$. Thus, we can use the asymptotic formula (\ref{a.6}) 
for the entire range of the left semicircle given in Eq.(\ref{a.5}).

The above discussion indicates that
it is appropriate to recast the $\Gamma(\alpha+\beta t_1)$ type 
in the integrand of Eq.(\ref{a.2}) into the $\Gamma(1-\alpha-\beta t_1)$ type by using the formula
 \begin{align}
\Gamma(\omega)\Gamma(1-\omega)=\frac{\pi}{\sin(\pi \omega)}
\label{a.7}
\end{align}
 in order to study the asymptotic behavior of $\log\abs{I}$.
One also needs the asymptotic behavior of the zeta function along the large left semicircle. 
 It may be convenient to use the formula
 \begin{align}
 \zeta(1-\omega)=2^{1-\omega}\pi^{-\omega}\zeta(\omega)\Gamma(\omega)\cos\left(\frac{\pi}{2}\omega\right).
\label{a.8}
 \end{align}
 By replacing $\omega$ by $1-\omega$ with $\omega =\alpha +\beta t_1$ in Eq.(\ref{a.8}), we have 
 \begin{align}
 \zeta(\alpha +\beta t_1)=\frac{(2\pi)^{\alpha +\beta t_1}}{\pi}
 \zeta(1-\alpha -\beta t_1)\Gamma(1-\alpha -\beta t_1)\sin\Bigl[\frac{\pi}{2}(\alpha +\beta t_1)\Bigr],
\label{a.9}
 \end{align}
 which will be used to recast the zeta function in Eq.(\ref{a.2}). It is easy to see 
 \begin{align}
 \log\abs{\sin\Bigl[\frac{\pi}{2}(\alpha +\beta t_1)\Bigr]}\simeq \frac{\pi}{2} \beta R \abs{\sin\varphi}
 \label{a.10}
 \end{align}
 and the asymptotic behavior of $\log\abs{\Gamma(1-\alpha -\beta t_1)}$ is obtained
 \footnote{Let us note that $\text{arg}~(1-\alpha-\beta t_1)\simeq \theta$ in the limit $R\rightarrow \infty$, where $\theta$ appears 
 in Eq.(\ref{a.6}).} by replacing $\alpha$ with $1-\alpha$ in Eq.(\ref{a.6}) in the limit $R\rightarrow \infty$.

Now we are ready to discuss the asymptotic behavior of $\log\abs{I}$. Let us recast 
the Gamma and zeta functions in Eq.(\ref{a.2}) by using the formulas
(\ref{a.7}) and (\ref{a.9}). Then, we obtain
\begin{align}
I=\pi (2\pi)^{2A}\frac{\zeta(1-2A-2t_1)\Gamma(1-2A-2t_1)}{\sin(\pi t_1)\Gamma(1-A-t_1)\Gamma(1-t_1)}
\left(\frac{ML_0}{4\pi}\right)^{-2t_1},
\label{a.11}
\end{align}
where we have denoted $A=\frac{1}{2}-t$. Aside from the irrelevant term, it follows that
\begin{align}
\log\abs{I}\simeq& \log\abs{\zeta(1-2A-2t_1)}+\log\abs{\Gamma(1-2A-2t_1)}-\log\abs{\sin(\pi t_1)}\notag\\
\hspace{5mm}&-\log\abs{\Gamma(1-A-t_1)}-\log\abs{\Gamma(1-t_1)}+\log\abs{\left(\frac{ML_0}{4\pi}\right)^{-2t_1}}.
\label{a.12}
\end{align}
We note that the argument of $\zeta(1-2A -2t_1)$
enters the region of absolute convergence as the $R$ tends to infinity, so that
it may be considered as a quantity of order one and thus, the first term in Eq.(\ref{a.12}) does not contribute 
to the asymptotic behavior.

Knowing that
 \begin{align}
 \log\abs{\left(\frac{ML_0}{4\pi}\right)^{-2t_1}}\simeq -2R\cos\varphi\log\left(\frac{ML_0}{4\pi}\right)
 \label{a.13}
 \end{align}
and using Eq.(\ref{a.6}), we finally obtain 
 \begin{align}
 \log\abs{I}\simeq -\pi R\abs{\sin\varphi} -2R\cos\varphi\log\left(\frac{ML_0}{2\pi}\right).
 \label{a.14}
 \end{align}
In order for the contribution to $\abs{I}$ from the large left semicircle $(\frac{\pi}{2}\leq \varphi\leq \frac{3}{2}\pi, R\rightarrow \infty)$ to 
be suppressed, one needs the condition
 \begin{align}
 0< \frac{ML_0}{2\pi} <1,
 \label{a.15}
 \end{align}
which is Eq.(\ref{3.5}) in the text.
%
%

Let us make a few comments in order. The first one is that the leading behavior $R\cos\varphi\log R$ and
the subleading one $R(\theta \sin\varphi +\cos\varphi)$ in Eq.({\ref{a.6}) cancel out between the 
three gamma functions in Eq.(\ref{a.12}). The dominant contribution is 
governed by the term linear in $R$, as shown in Eq.(\ref{a.14}). The second one is that 
the complex integral in Eq.(\ref{3.12}) is essentially the same type as Eq.(\ref{a.1}), so that
a similar discussion leads to the condition (\ref{3.11}). The last one is that if
we take $\varphi=\tfrac{\pi}{2}~(\tfrac{3}{2}\pi)$ in Eq.(\ref{a.14}), we have $\log\abs{I}\simeq -\pi R$.
This implies that the contour integral  (\ref{a.1}) (and (\ref{b.1}) defined below) along the imaginary axis 
is well-defined, as it should be. The three comments are also applicable to the case of the fermion, which is given below.

Let us next derive the condition (\ref{4.5}) in the case of the fermion. The relevant 
part of the $t_1$-integral in Eq.(\ref{4.4}) is given by
\begin{align}
\int_{c_1-i\infty}^{c_1+i\infty}dt_1~\Gamma(t_1-t+\tfrac{1}{2})\eta(2t_1-2t+1)\Gamma(t_1)\left(\frac{ML_0}{2}\right)^{-2t_1}.
\label{a.16}
\end{align}
The only difference, compared with Eq.(\ref{a.1}), is that the zeta function in Eq.(\ref{a.1}) is replaced by 
the eta function. We denote the integrand in Eq.(\ref{a.16}) by $I_f$.

Using the formula (\ref{a.7}) and (\ref{4.3}) together with (\ref{a.9}), the integrand $I$ is recast as
\begin{align}
 I_f=\pi (2\pi)^{2A} \frac{(1-2^{1-2A-2t_1})\zeta(1-2A-2t_1)\Gamma(1-2A-2t_1)}{\sin(\pi t_1)\Gamma(1-A-t_1)\Gamma(1-t_1)}
 \left(\frac{ML_0}{4\pi}\right)^{-2t_1}.
  \label{a.17}
 \end{align}
 Up to the irrelevant terms, we have
 \begin{align}
 \log\abs{I_f}\simeq &\log\abs{(1-2^{1-2A-2t_1})}+\log\abs{\zeta(1-2A-2t_1)}+\log\abs{\Gamma(1-2A-2t_1)}\notag\\
 \hspace{5mm}&-\log\abs{\sin(\pi t_1)} -\log\abs{\Gamma(1-A-t_1)}-\log\abs{\Gamma(1-t_1)}
 +\log\abs{\left(\frac{ML_0}{4\pi}\right)^{-2t_1}}.
 \label{a.18}
 \end{align}
The first term in Eq.(\ref{a.18}) behaves like
\begin{align}
\log\abs{(1-2^{1-2A-2 t_1 })}\simeq -2 R \cos\varphi \log 2
\label{a.19}
\end{align}
in the limit $R\rightarrow \infty$ and the second term can be ignored as explained before. We immediately obtain
 \begin{align}
\log\abs{I_f}\simeq -\pi R\abs{\sin\varphi}-2R\cos\varphi\log\left(\frac{ML_0}{\pi}\right),
 \label{a.20}
\end{align}
where we have used Eqs.(\ref{a.12}), (\ref{a.14}) and (\ref{a.19}). 
The difference of the factor $2$ in the logarithm of the second term in Eq.(\ref{a.20}), compared with Eq.(\ref{a.14}), comes from 
the difference between the zeta function and the eta function (see Eqs.(\ref{4.3}) and (\ref{a.19})). In order for 
the contribution to $\abs{I_f}$ from the large left semicircle $(\frac{\pi}{2}\leq \varphi\leq \frac{3}{2}\pi, R\rightarrow\infty)$ to 
be suppressed, one obtains the condition
 \begin{align}
 0< \frac{ML_0}{\pi} < 1,
 \label{a.21}
 \end{align}
 which is (\ref{4.5}) in the text.
%
%
%
%
%
%

It may be interesting to note that the antiperiodic boundary condition for the Euclidean time direction for the 
fermion can be viewed as the periodic one if we consider $2L_0$. This suggests that the condition for 
the scalar case (\ref{a.15}) is reduced to the fermion one (\ref{a.21}) by replacing $L_0$ by $2L_0$ in 
Eq.(\ref{a.15}), which is actually the case.

It is worth noting the following point. Our results (\ref{3.13}), (\ref{3.16}), (\ref{4.14}) and (\ref{4.17})
contain the infinite series. One can show that the conditions (\ref{a.15}) and (\ref{a.21}), which have 
been imposed to obtain the results and are derived in this Appendix, exactly 
coincide with the radius of convergence for the series. This fact strongly supports the validity of the results
using the framework to study the nonanalytic terms in this paper.

%
%
%
%
\section{Residue evaluation for $F_{L_k}^{(1)D-k}(M)$}
Let us derive the results (\ref{3.13}) and (\ref{3.16}) in this Appendix. We 
consider the complex integral (\ref{3.12})
\begin{align}
F_{L_k}^{(1)D-k}(M)&
&=\frac{-2}{(2\pi)^{\frac{D-k}{2}}}\left(\frac{M^2}{2}\right)^{\frac{D-k}{2}}
\frac{1}{4\pi i} \int_{c-i\infty}^{c+i\infty}dt~\Gamma(t-\tfrac{D-k}{2})\Gamma(t)\zeta(2t)\left(\frac{ML_k}{2}\right)^{-2t}.
\label{b.1}
\end{align}
We deform the contour in the $t$ plane so that it encloses all the poles of the integrand 
and evaluate the $t$ integral by the residue theorem under the condition (\ref{3.11}).
Note that the integrand (\ref{b.1}) can have at most a double pole because the
part $\Gamma(t)\zeta(2t)$ obviously never possesses the double pole. Thus, the double pole 
exists in the part $\Gamma(t-\tfrac{D-k}{2})\Gamma(t)$ or $\Gamma(t-\tfrac{D-k}{2})\zeta(2t)$
in the integrand, depending on whether $\tfrac{D-k}{2}$ is even or odd. It is convenient 
to evaluate the integral by separating the cases in which $D$ and $k$ are even or odd.

We first present the necessary formulas to evaluate the integral using residues. Let us consider 
the function $F(t)=A(t)B(t)C(t)$ and suppose that $t=\alpha$ is a double pole of the 
part $A(t)B(t)$ in $F(t)$ and $C(t)$ is regular at the pole. Writing 
\begin{align}
A(t)&=\frac{A_{\text{res}}}{t-\alpha}+A_{\text f}+O(t-\alpha)
\label{b.2}
\\
B(t)&=\frac{B_{\text{res}}}{t-\alpha}+B_{\text f}+O(t-\alpha),
\label{b.3}
\end{align}
the residue of $F(t)$ at the pole $t=\alpha$ is calculated to be 
\begin{align}
{\text {Res}}~F(t)\Big|_{t=\alpha}&=\frac{d}{dt}\left(F(t)(t-\alpha)^2\right)\Big|_{t=\alpha}\notag\\
&=A_{\text{res}}B_{\text{res}}C'(\alpha)+A_{\text{res}}B_{\text{f}}C(\alpha)+A_{\text{f}}B_{\text{res}}C(\alpha).
\label{b.4}
\end{align}
The $C'(\alpha)$ in the first term of Eq.(\ref{b.4}) is the origin of the logarithmic-type nonanalytic 
terms, as we will see below. If the pole $t=\alpha$ is the trivial zero of $C(t)$, that is, $C(\alpha)=0$, then, we have
\begin{align}
F(t)=\frac{A_{\text{res}}B_{\text{res}}}{t-\alpha}C'(\alpha)+O(t-\alpha)\quad\mbox{for}\quad C(\alpha)=0,
\label{b.5}
\end{align}
so that the pole $t=\alpha$ becomes a single pole of the integrand to yield
\begin{align}
{\text {Res}}~F(t)\Big|_{t=\alpha}=A_{\text{res}}B_{\text{res}}C'(\alpha)\quad\mbox{for}\quad C(\alpha)=0.
\label{b.6}
\end{align}
Eqs.(\ref{b.4}) and  (\ref{b.6}) are frequently used in evaluating the integral by residues.
\subsection{$(D, k)=(\text{even}, \text{even}), (\text{odd}, \text{odd})$}
The first term in Eq.(\ref{3.13}) arises from the the single pole $t=\tfrac{1}{2}$ of $\zeta(2t)$
in Eq.(\ref{3.12}) and is the power-type nonanalytic term that we have already obtained in Eq.(\ref{4.10}) of the paper II.

The part $\Gamma(t-\tfrac{D-k}{2})\Gamma(t)$ possesses a double pole at $t=0$ in this case.
By using Eq.(\ref{b.4}) and denoting $\ell_k\equiv\tfrac{D-k}{2}$, the residue at the pole of Eq.(\ref{3.12}) is calculated to be
\begin{align}
&\frac{-2}{(2\pi)^{\frac{D-k}{2}}}\left(\frac{M^2}{2}\right)^{\frac{D-k}{2}}\frac{1}{2}
\Bigg\{
\frac{(-1)^{\ell_k}}{\ell_k !}\left(\zeta(2t)\left(\frac{ML_k}{2}\right)^{-2t}\right)'\Bigg|_{t=0}+
\frac{(-1)^{\ell_k}}{\ell_k !}\psi(1) \zeta(0)\notag\\
&\hspace{10mm}
+\frac{(-1)^{\ell_k}}{\ell_k !}\psi(\ell_k+1)\zeta(0)
\Biggr\}\notag\\
&=\frac{-1}{2^{D-k}\pi^{\frac{D-k}{2}}}
\frac{(-1)^{\frac{D-k}{2}}}{(\frac{D-k}{2})!}
M^{D-k}\Bigg\{
\log\left(\frac{ML_k}{4\pi}\right)-\frac{1}{2}\left(\psi(1) +\psi(\tfrac{D-k}{2} +1)\right)
\Biggr\},
\label{b.7}
\end{align}
where we have used the expansion
\begin{align}
\Gamma(t)=\frac{(-1)^n}{n!}\frac{1}{t+n}+\frac{(-1)^n}{n!}\psi(n+1)+O(t+n).
\label{b.8}
\end{align}
Here, the digamma function $\psi(m)$ is given by
\begin{align}
\psi(m)=\sum_{r=1}^{m-1}\frac{1}{r}-\gamma_E\qquad \mbox{and}\qquad \psi(1)=-\gamma_E .
\label{b.9}
\end{align}
We have also used $\zeta(0)=-\tfrac{1}{2}$ and $\zeta'(0)=-\tfrac{1}{2}\log(2\pi)$ in the 
calculation and $\gamma_E$ is the Euler's constant. Eq.(\ref{b.7}) is the 
second term in Eq.(\ref{3.13}). The logarithmic-type nonanalytic term comes from 
$C'(\alpha)$ of Eq.(\ref{b.4}) in evaluating the residue at the double pole.

Let us next compute the contribution from the single pole $t=-\ell~(\ell=1,2,\cdots)$ of
$\Gamma(t)$ and $\Gamma(t-\tfrac{D-k}{2})$, which corresponds to the case (\ref{b.6}) 
because of $\zeta(-2\ell)=0~(\ell=1,2,\cdots)$. The straightforward calculation shows
\begin{align}
&\frac{-2}{(2\pi)^{\frac{D-k}{2}}}\left(\frac{M^2}{2}\right)^{\frac{D-k}{2}}\frac{1}{2}
\sum_{\ell =1}^{\infty}\frac{(-1)^{\ell+\ell_k}}{(\ell+\ell_k)!}\frac{(-1)^{\ell}}{\ell!}2 \zeta'(-2\ell)\left(\frac{ML_k}{2}\right)^{2\ell}
\notag\\
&\hspace{20mm}=\frac{(-1)(-1)^{\frac{D-k}{2}}}{2^{D-k}\pi^{\frac{D-k}{2}}}M^{D-k}\sum_{\ell=1}^{\infty}
\frac{(-1)^{\ell}\Gamma(2\ell +1)\zeta(2\ell +1)}{\Gamma(\ell+\tfrac{D-k}{2}+1)\Gamma(\ell+1)}
\left(\frac{ML_k}{4\pi}\right)^{2\ell},
\label{b.10}
\end{align}
where we have used 
\begin{align}
\zeta'(-2\ell)=\frac{(-1)^{\ell}(2\ell)!}{2(2\pi)^{2\ell}}\zeta(2\ell +1)=
\frac{(-1)^{\ell}\Gamma(2\ell +1)}{2(2\pi)^{2\ell}}\zeta(2\ell +1).
\label{b.11}
\end{align}
Eq.(\ref{b.10}) is the third term in Eq.(\ref{3.13}).

We finally calculate the contribution from the the single 
pole $t=\ell_k-m=\tfrac{D-k}{2}-m~(m=0,1,\cdots, \tfrac{D-k}{2}-1)$ of $\Gamma(t-\tfrac{D-k}{2})$. We find 
\begin{align}
&\frac{-2}{(2\pi)^{\frac{D-k}{2}}}\left(\frac{M^2}{2}\right)^{\frac{D-k}{2}}\frac{1}{2}
\sum_{m=0}^{\ell_k -1}\frac{(-1)^{m}}{m!}\Gamma(\ell_k- m)\zeta(2\ell_k-2m)\left(\frac{ML_k}{2}\right)^{-2(\ell_k -m)}
\notag\\
&\hspace{15mm}=\frac{-1}{\pi^{\frac{D-k}{2}}L_k^{D-k}}\sum_{m=0}^{\frac{D-k}{2}-1}
\frac{(-1)^m}{m!}\Gamma(\tfrac{D-k}{2}-m)\zeta(D-k -2m)
\left(\frac{ML_k}{2}\right)^{2m}.
\label{b.12}
\end{align}
We have obtained the fourth term and have completed the computation of the four terms in Eq.(\ref{3.13}). 
\subsection{$(D, k)=(\text{even}, \text{odd}), (\text{odd}, \text{even})$}
The first term in Eq.(\ref{3.16}) arises from the single pole $t=0$ of $\Gamma(t)$ and 
is the power-type nonanalytic term that we have already obtained in the paper II.

The part $\Gamma(t-\tfrac{D-k}{2})\zeta(2t)$ has the double pole $t=\tfrac{1}{2}$ in this case. We 
evaluate the residue at the pole and the result is 
given, denoting $m_k=\tfrac{D-k-1}{2}$, by
\begin{align}
&\frac{-2}{(2\pi)^{\frac{D-k}{2}}}\left(\frac{M^2}{2}\right)^{\frac{D-k}{2}}\frac{1}{2}
\Biggl\{
\frac{(-1)^{m_k}}{m_k!}\frac{1}{2}\left(\Gamma(t)\left(\frac{ML_k}{2}\right)^{-2t}\right)'\Bigg|_{t=\frac{1}{2}}
+\frac{(-1)^{m_k}}{m_k!}\gamma_E \Gamma(\tfrac{1}{2})\left(\frac{ML_k}{2}\right)^{-1}\notag\\
&\hspace{5mm}+\frac{(-1)^{m_k}}{m_k!}\psi(m_k+1)\frac{1}{2}\Gamma(\tfrac{1}{2})\left(\frac{ML_k}{2}\right)^{-1}
\Biggr\}\notag\\
&=\frac{-1}{2^{D-k-1}\pi^{\frac{D-k-1}{2}}}\frac{(-1)^{\frac{D-k-1}{2}}}{(\frac{D-k-1}{2})!}\frac{M^{D-k-1}}{L_k}
\Bigl\{
-\log\left(ML_k\right)+\frac{1}{2}\left(\gamma_E+\psi(\tfrac{D-k-1}{2}+1)\right)
\Bigr\}.
\label{b.13}
\end{align}
This is the second term in Eq.(\ref{3.16}). In the calculation we have used 
$\Gamma(\tfrac{1}{2})=\sqrt{\pi}$ and $\Gamma'(\tfrac{1}{2})=\Gamma(\tfrac{1}{2})\psi(\tfrac{1}{2})
=\sqrt{\pi}(-\log 4 -\gamma_E)$. The Euler's constant, $\gamma_E$ in Eq.(\ref{b.13}) comes from $\Gamma'(\frac{1}{2})$ and 
the finite term of the expansion
\begin{align}
\zeta(2t)=\frac{1}{2}\frac{1}{t-\frac{1}{2}}+\gamma_E+O(t-\tfrac{1}{2}).
\label{b.14}
\end{align}
The logarithmic-type nonanalytic term arises from the $C'(\alpha)$ of Eq.(\ref{b.4}) in evaluating the
residue at the double pole.

Let us next study the contribution from the single pole $t=\tfrac{D-k}{2}-m$
$(m=0,1,\cdots)$, excluding $m=\tfrac{D-k-1}{2}$, of $\Gamma(t-\tfrac{D-k}{2})$ and it is given by
\begin{align}
&\frac{-2}{(2\pi)^{\frac{D-k}{2}}}\left(\frac{M^2}{2}\right)^{\frac{D-k}{2}}\frac{1}{2}
\sum_{\substack{m=0\\(m\neq \frac{D-k-1}{2})}}^{\infty}\frac{(-1)^m}{m!}\Gamma(\tfrac{D-k}{2}-m)\zeta(D-k-2m)
\left(\frac{ML_k}{2}\right)^{-2(\frac{D-k}{2}-m)}\notag\\
&=\frac{-1}{\pi^{\frac{D-k}{2}}L_k^{D-k}}\sum_{\substack{m=0\\(m\neq \frac{D-k-1}{2})}}^{\infty}\frac{(-1)^m}{m!}
\Gamma(\tfrac{D-k}{2}-m)\zeta(D-k -2m)\left(\frac{ML_k}{2}\right)^{2m}.
\label{b.15}
\end{align}
This is the third term and we have obtained the three terms in Eq.(\ref{3.16}). 
%
%
%
%
%
%
\section{Residue evaluation for $F_{L_0}^{(1)(D, f)}(M)$}
Let us obtain the nonanalytic terms in the effective potential for the fermion in section $4$. 
To this end, we need to compute the integral in the second term of Eq.(\ref{4.10}) 
\begin{align}
F_{L_0}^{(1)(D,f)}(M)=
-{\cal N}\frac{1}{L_0^D\pi^{\frac{D}{2}}}\frac{1}{2\pi i}\int_{c-i\infty}^{c+i\infty}
dt~\Gamma(t+\tfrac{D}{2})\eta(2t+D)\Gamma(t)\left(\frac{ML_0}{2}\right)^{-2t},
\label{c.1}
\end{align}
where we have rewritten $t_1$ and $c_1$ as $t$ and $c$, respectively.
We deform the contour in the $t$ plane so as to encircle all the poles of the integrand 
and evaluate the $t$ integral by the residue theorem under the condition (\ref{a.16}).
Let us note that the eta function has no pole unlike the zeta function in Eq.(\ref{b.1}).
We consider the integral separately, depending on whether $D$ is even or odd.
\subsection{$D=$ even}
The double pole exists in the part $\Gamma(t+\tfrac{D}{2})\Gamma(t)$ in this case
and taking into account $\eta(-2m)=0~(m=1,2,\cdots)$, it is given by $t=-\tfrac{D}{2}$.
From (\ref{b.4}), we find
\begin{align}
&-{\cal N}\frac{1}{\pi^{\frac{D}{2}}L_0^D}\Biggl\{
\frac{(-1)^{\frac{D}{2}}}{(\frac{D}{2})!}\left(\eta(2t+D)\left(\frac{ML_0}{2}\right)^{-2t}\right)'\Bigg|_{t=-\frac{D}{2}}
+\frac{(-1)^{\frac{D}{2}}}{(\frac{D}{2})!}\psi(\tfrac{D}{2}+1)\eta(0)\left(\frac{ML_0}{2}\right)^D\notag\\
&\hspace{5mm}+\psi(1)\frac{(-1)^{\frac{D}{2}}}{(\frac{D}{2})!}\eta(0)\left(\frac{ML_0}{2}\right)^D
\Biggr\}\notag\\
&={\cal N}\frac{(-1)^{\frac{D}{2}}}{2^D\pi^{\frac{D}{2}}\left(\frac{D}{2}\right)!}
M^D 
\Biggl\{\log\left(\frac{ML_0}{\pi}\right)-\frac{1}{2}\left(\psi(1)+\psi(\tfrac{D}{2}+1)\right)
\Biggr\},
\label{c.2}
\end{align}
where we have used $\eta(0)=\tfrac{1}{2}, \eta'(0)=\tfrac{1}{2}\log(\tfrac{\pi}{2})$. This is the first term in Eq.(\ref{4.14}).
We see that the logarithmic-type nonanalytic term comes from $C'(\alpha)$ of Eq.(\ref{b.4}) in evaluating 
the residue at the double pole.

Let us next study the contribution from the single pole $t=-\tfrac{D}{2}-\ell~(\ell=1,2,\cdots)$ of $\Gamma(\tfrac{D}{2}+t)$
and $\Gamma(t)$, which is the case for Eq.(\ref{b.6}) because of $\eta(-2\ell)=0~(\ell=1,2,\cdots)$. We have
\begin{align}
&-{\cal N}\frac{1}{\pi^{\frac{D}{2}}L_0^D}\sum_{\ell=1}^{\infty}\frac{(-1)^{\ell}}{\ell !}\frac{(-1)^{\ell +\frac{D}{2}}}{(\ell +\frac{D}{2})!}
2\eta'(-2\ell)\left(\frac{ML_0}{2}\right)^{-2(-\frac{D}{2}-\ell)}\notag\\
&\hspace{20mm}=-{\cal N}\frac{(-1)^{\frac{D}{2}}}{2^D\pi^{\frac{D}{2}}}M^D
\sum_{\ell=1}^{\infty}\frac{(-1)^{\ell}\Gamma(2\ell +1)(1-2^{1+2\ell})}{\Gamma(\ell+\tfrac{D}{2}+1)\Gamma(\ell+1)}
\zeta(2\ell +1)\left(\frac{ML_0}{4\pi}\right)^{2\ell},
\label{c.3}
\end{align}
where we have used 
\begin{align}
\eta'(-2\ell)=(1-2^{1+2\ell})\zeta'(-2\ell)=(1-2^{1+2\ell})
\frac{(-1)^{\ell}\Gamma(2\ell +1)}{2(2\pi)^{2\ell}}\zeta(2\ell +1)
\label{c.4}
\end{align}
and Eq.(\ref{b.11}). This is the second term in Eq.(\ref{4.14}).

Let us finally calculate the contribution from the single pole $t=-m~(m=0,1,\cdots, \tfrac{D}{2}-1)$ of $\Gamma(t)$ and we obtain
\begin{align}
-{\cal N}\frac{1}{\pi^{\frac{D}{2}}L_0^D}\sum_{m=0}^{\frac{D}{2}-1}\frac{(-1)^{m}}{m !}\Gamma(\tfrac{D}{2}-m)
\eta(D-2m)\left(\frac{ML_0}{2}\right)^{2m},
\label{c.5}
\end{align}
which is the third term and we have finished computing the three terms in Eq.(\ref{4.14}). 
\subsection{$D=$ odd}
The integrand possesses no double pole in this case, so that there exists no logarithmic-type nonanalytic term. 
The contributions from the single pole $t=-\tfrac{D}{2}$ of $\Gamma(t+\tfrac{D}{2})$
and the one $t=-\ell~(\ell=0,1,\cdots)$ of $\Gamma(t)$ yield\footnote{One might think that the single poles
$t=-\frac{D}{2}-m~(m=1,2,\cdots)$ of $\Gamma(t+\tfrac{D}{2})$ contribute to Eq.(\ref{c.1}). This is not the case 
because the combination $\Gamma(t+\tfrac{D}{2})\eta(2t+D)$ has no pole there due to the property 
$\eta(-2m)=0~(m=1,2,\cdots)$.}
\begin{align}
%
%
-{\cal N}\frac{(-1)^{\frac{D+1}{2}}}{2^{\frac{D+1}{2}}\pi^{\frac{D-1}{2}}D!!}M^D
-{\cal N}\frac{1}{\pi^{\frac{D}{2}}L_0^D}\sum_{\ell=0}^{\infty}\frac{(-1)^{\ell}}{\ell !}\Gamma(\tfrac{D}{2}-\ell)
\eta(D-2\ell)\left(\frac{ML_0}{2}\right)^{2\ell}.
\label{c.6}
\end{align}
The first term is explicitly given by Eq.(\ref{2.18}) with $f=1$ and an extra minus sign. We 
have calculated the two terms in Eq.(\ref{4.17}). 
%
%
%
%

\end{document}